\documentclass[%
 reprint,
superscriptaddress,
 amsmath,amssymb,
 aps,
prb,
]{revtex4-1}

\usepackage{graphicx}
\usepackage{epstopdf}
\usepackage{dcolumn}
\usepackage{bm}
\usepackage{gensymb}
\usepackage{upgreek}
\usepackage{hyperref}

\begin{document}

\preprint{APS/123-QED}

\title{Enhancement of Superconductivity in WP via Oxide-Assisted Chemical Vapor Transport}

\author{Daniel J. Campbell}
\thanks{campbell94@llnl.gov}
\affiliation{Maryland Quantum Materials Center, Department of Physics, University of Maryland, College Park, Maryland 20742, USA}
\affiliation{Physics Division, Lawrence Livermore National Laboratory, 7000 East Avenue, Livermore, CA 94550 USA}
\author{Wen-Chen Lin}
\author{John Collini}
\author{Yun Suk Eo}
\author{Yash Anand}
\author{Shanta Saha}
\affiliation{Maryland Quantum Materials Center, Department of Physics, University of Maryland, College Park, Maryland 20742, USA}
\author{Dave Graf}
\affiliation{National High Magnetic Field Laboratory, 1800 East Paul Dirac Drive, Tallahassee, Florida 32310, USA}
\author{Peter Y. Zavalij}
\affiliation{Department of Chemistry, University of Maryland, College Park, Maryland 20742, USA}
\author{Johnpierre Paglione}
\thanks{paglione@umd.edu}
\affiliation{Maryland Quantum Materials Center, Department of Physics, University of Maryland, College Park, Maryland 20742, USA}
\affiliation{Canadian Institute for Advanced Research, Toronto, Ontario M5G 1Z8, Canada}

\date{\today}

\begin{abstract}

Tungsten monophosphide (WP) has been reported to superconduct below 0.8~K, and theoretical work has predicted an unconventional Cooper pairing mechanism. Here we present data for WP single crystals grown by means of chemical vapor transport (CVT) of WO$_3$, P, and I$_2$. In comparison to synthesis using WP powder as a starting material, this technique results in samples with substantially decreased low-temperature scattering and favors a more three dimensional morphology. We also find that the resistive superconducting transitions in these samples begin above 1~K. Variation in $T_c$ is often found in strongly correlated superconductors, and its presence in WP could be the result of influence from a competing order and/or a non $s$-wave gap.

\end{abstract}

\maketitle

\section{Introduction}

Though first investigated half a century ago~\cite{RundqvistPhosphides,RundqvistWP,BellavanceFePPreparation}, the MnP (or B31) type transition metal pnictide binaries have been the subject of renewed attention in recent years. The various members show a wide variety of properties of interest in condensed matter physics: unusual spin density wave ordering~\cite{RodriguezFeAsSDW}, doping- and pressure-induced superconductivity~\cite{HiraiRuRhPn,WuCrAs,KotegawaCrAs,ChengMnP}, and topological band structure features~\cite{NiuCrAs,CuonoMnPType,CampbellFeP}. Recently, superconductivity was found in WP at 0.8~K~\cite{LiuWP}. Despite the low reported $T_c$ and $H_{c2}$ ($<$~20~mT), the superconducting state in WP attracts attention because it is predicted to be topologically nontrivial, like those observed in isostructural CrAs and MnP under pressure~\cite{CuonoMnPType,NigroWP}. This is thought to be a result of nonsymmorphic crystallographic symmetries, which protect topologically nontrivial elements of the band structure~\cite{CuonoMnPType}. The fact that superconductivity is present in WP without the need to apply pressure makes it the most appealing of the three materials for continued study and application.

Producing WP requires combining W, which has the highest melting point of any metal, with a very high vapor pressure element in P. It can therefore be challenging to find an appropriate temperature for crystal growth. Here, we sidestep the difficulty of a direct reaction between these two elements by using WO$_3$ for single crystal synthesis. Single crystals grown through chemical vapor transport of WO$_3$, P, and I$_2$ (as the transport agent) are typically more three dimensional than the long, thin needles reported in previous work~\cite{LiuWP,NigroWP}. These samples also have a lower residual resistivity, a sign of reduced low temperature scattering. With this comes a large magnetoresistance (MR) that is linear in one orientation up to 20~T, another property that has been associated with topological band structure points in isostructural CrAs~\cite{NiuCrAs} and FeP~\cite{CampbellFeP}. Most notably of all, the onset of superconductivity in transport measurements comes at temperatures as high as 1.3~K in the oxide-grown samples. A disorder-dependent $T_c$ is an indicator of unconventional superconductivity~\cite{MackenzieSr2RuO4}, and the unusual broadening of the transition at higher temperature that we see occurs in other unconventional superconductors as a result of competing orders~\cite{ParkCeRhIn5TcVariation,ChenLaNiC2,LiuSnSb}. The discovery via x-ray of multiple crystal modulation vectors could be related to this. The technique with which WP is grown has a clear influence on its properties. Here we outline our procedure and go through the measured properties, including those that are only now apparent with the obtained three dimensional geometry.

\section{Methods}
Transport measurements were performed in a Quantum Design Physical Properties Measurement System, where temperatures below 1.8~K were achieved with either a $^3$He refrigerator or an adiabatic demagnetization refrigerator (ADR). The magnetic field was oscillated to zero to remove any trapped flux before measurement. High magnetic field measurements were done with a 41.5~T resistive magnet and $^3$He system at the National High Magnetic Field Laboratory (NHMFL). Magnetic susceptibility was measured with a 7~T Quantum Design SQUID-VSM Magnetic Properties Measurement System. Powder x-ray diffraction (XRD) measurements were performed with a Rigaku Miniflex diffractometer using Cu $K_{\alpha{}}$ radiation. Single crystal diffraction was done with a Bruker Smart Apex II CCD diffractometer on two crystals at temperatures between 120 and 300~K. The integral intensity were correct for absorption using SADABS software~\cite{KrauseStructure} using multi-scan method. Resulting minimum and maximum transmission are 0.033 and 0.082 respectively. The structure was solved with the ShelXT (Sheldrick, 2015a)~\cite{SheldrickShelXT} program and refined with the ShelXL~\cite{SheldrickShelXL} program and least-square minimisation using ShelX software package~\cite{SheldrickShelXL}. Single crystals were aligned using both the Miniflex and Laue photography, and further elementally characterized with energy-dispersive x-ray spectroscopy (EDS).

\section{Single Crystal Synthesis}

Chemical vapor transport (CVT) has been used to grow high quality crystals of practically all other phosphides in the MnP-type structure family~\cite{BinnewiesCVTBook,BellavanceFePPreparation,NiuCrP}. For this technique, reactants are placed at one end of an evacuated quartz ampule with a temperature gradient along its length. This temperature gradient results in different solid-gas reaction rates at the two ends of the ampule, and if done correctly favors crystallization of the desired material at the ampule end that was originally empty. Iodine acts as a transport agent, whose high vapor pressure facilitates the evaporation and movement of the other materials down the length of the tube. The high vapor pressure of phosphorus means that many P-containing materials can be readily synthesized with this technique.

In the only reports on superconducting WP, crystals were grown via CVT with prereacted WP powder and iodine~\cite{LiuWP, NigroWP,ZhangWPRamanTheory}. However, in the case of WP this procedure has several possible pitfalls. Our work with FeP has shown that samples had a much lower residual resistivity in CVT growths starting from the elements compared to when prereacted FeP powder was used~\cite{CampbellFeP}. The residual resistivity is the baseline, temperature-independent value seen at the lowest temperatures, and comes primarily from inherent impurity scattering. This means that lower values generally signify a smaller impurity concentration.

Another concern is the vastly different behavior of the two elements at high temperature. Red phosphorus (the least volatile form of the element, and thus the one used for crystal growth) sublimes below 600~$\degree$C, while tungsten does not melt until 3400~$\degree$C and has one of the lowest vapor pressures of any element. It is difficult to find a middle point between these two for a reaction, especially given that in practice we are also limited by the fact that quartz ampules will soften or melt above about 1250~\degree{}C; there are, in fact, few examples of CVT involving pure W~\cite{LenzCVTDevelopments}. For that reason we took inspiration from an earlier work that used the oxides of heavy transition metals (W, Hf, Ta, and others) to produce single crystal transition metal phosphides with CVT~\cite{MartinOxidesCVT}, and which was employed in later work to produce WP as well as WP$_2$~\cite{MathisWO3PReduction,SchoenemannWP2}. WO$_3$ has a much lower melting point (1473~\degree{}C) than W; it is also thought that the release of oxygen at high temperature in combination with I$_2$ is beneficial to gas phase transport~\cite{MathisWO3PReduction,LenzCVTDevelopments}. Another group has reported using extremely high temperatures and pressures (3200~\degree{}C and 5~GPa) to achieve a congruent W-P melt~\cite{XiangWPGrowth} and produce large crystals. However, the need for significantly higher temperature and pressure makes such synthesis technically challenging.

\begin{figure}
    \centering
    \includegraphics[width=0.48\textwidth]{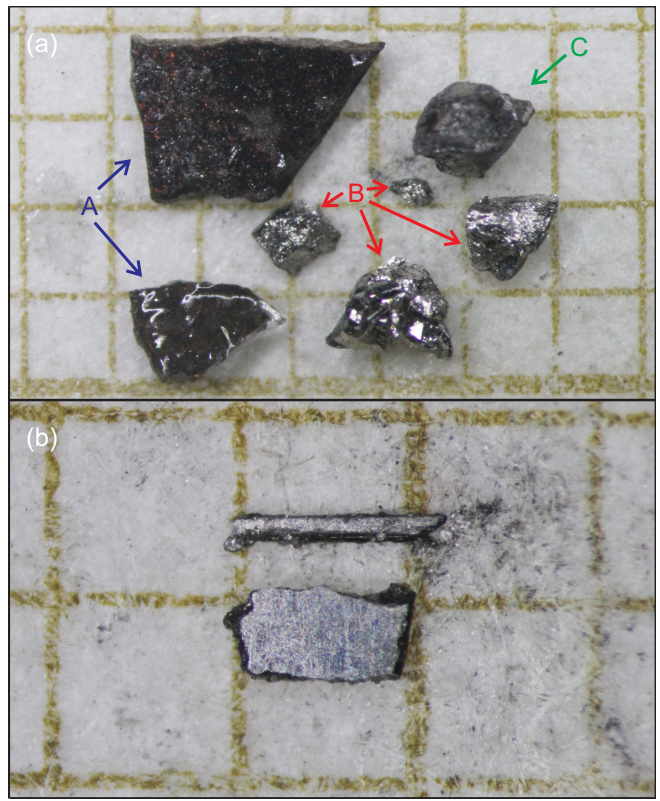}
    \caption{(a) The different materials that form at the cold end of a sealed quartz ampule when starting with WO$_3$, P, and I$_2$ at the hot end. Pieces marked A are chunks of W$_8$P$_4$O$_{32}$, those marked B are WP (either single crystals, or multiple intergrown crystals), and C is a piece of WP fused to a piece of W$_8$P$_4$O$_{32}$. (b) A comparison of a (top) needlelike and (bottom) platelike crystal of WP, after initial polishing of only a single pair of opposing faces. The latter morphology was much more common in oxide-assisted growths. The grid paper in both pictures is composed of 1~$\times{}$~1~mm$^2$ squares.}
    \label{fig:WPCrystals}
\end{figure} 

For our CVT growths, WO$_3$ (CERAC, 99.9\% pure) and red P (Sigma Aldrich, $>$99.99\% trace metals basis) powders were ground together in a 1:1 ratio and placed into a quartz ampule, with additional I$_2$ (J.T. Baker, 99.9\%, about 1~mg/cm$^3$). The ampule was half the length of a single zone tube furnace (about 15~cm), and oriented so that the reactants were initially in the middle of a furnace, at temperatures in the range 900-1000~\degree{}C. The ampule end at the edge of the furnace was about 200~\degree{}C colder, and the growth was left for 10-14 days. Afterwards, it was found that some powder remained at the hot end, while a mixture of powder and crystalline material was at the cold end [Fig.~\ref{fig:WPCrystals}(a)]. XRD of the hot end powder showed that it was pure WP, while the cold end was a mixture of WP powder and single crystals, which were black and shiny, with larger, red-tinged chunks of W$_8$P$_4$O$_{32}$ (equivalently, W$_2$O$_4$[PO$_4$]). Often the WP crystals were found fused to the W$_8$P$_4$O$_{32}$, but the combination could be polished to leave just the binary [Fig.~\ref{fig:WPCrystals}(b), lower sample], with the absence of the other phase confirmed by XRD, EDS, and low temperature measurements. The crystals produced in this growth are much more three dimensional than the needles reported with WP powder-based growth~\cite{LiuWP,NigroWP}, though a small number of needlelike crystals were found in oxide-based growths as well [Fig.~\ref{fig:WPCrystals}(b), upper sample]. Subsequent attempts to use the WP powder that remained at the hot end for a new CVT growth with I$_2$ and a similar temperature and time profile to the oxide growth resulted in the powder staying in the hot region, without transporting to the cold ampule end. This is further evidence that oxygen is a critical part of the transport and crystal nucleation processes.

\section{Normal State Properties}

\begin{table*}
	\centering	
\caption{Atomic position and and anisotropic displacement parameters (in units of \AA$^2$) for a WP single crystal at 120~K. U$_{12}$ and U$_{23}$ are both zero. } 
\setlength{\tabcolsep}{6pt}
\begin{tabular}{ l c c c c c c c r }
\hline \hline									
Atom		&	$x$ &		$y$	&	$z$ 	& U$_{eq}$ &  U$_{11}$ & U$_{22}$	& U$_{33}$  & U$_{13}$ \\
\hline																		\\
W					&	0.51323(3)			&	0.25				&	0.68851(3)	& 0.00125(8)			&			0.0074(10)		&	0.00195(10) &	0.00106(10)	& -0.0006(4)	\\ 
P					&	0.18441(17)		&		0.25			&	0.43383(18)	& 0.00205(18)				&			0.0022(4)		&	0.0021(4)    &   0.0019(4) & -0.0004(3)			\\ \hline\hline
\end{tabular}
\label{XRDPositions}
\end{table*}

A 3D single crystal was chosen from one of the growths for detailed XRD measurements at 120~K. It was confirmed to be in the MnP-type $Pnma$ orthorhombic structure with lattice parameters  \textit{a}~=~5.7322(5)~\AA{}, \textit{b}~=~3.2487(3)~\AA{}, and \textit{c}~=~6.2246(5)~\AA{}, quite similar to previous reports at ambient temperature~\cite{MartinOxidesCVT,LiuWP,TayranWPTheory} in spite of possible thermal expansion effects. Further results from the structural refinement are given in Table~\ref{XRDPositions}. XRD was also performed on a crystal grown with the oxide method that had a more needlelike shape (such as the upper sample in Fig.~\ref{fig:WPCrystals}(b)) with very similar results. The x-ray measurements picked out a superstructure modulation vector of (0 $\frac{1}{7}$~$\frac{1}{7}$) in both samples, as well as additional vectors of (0 $\frac{1}{2}~\frac{1}{2}$), ($\frac{1}{2}~0~\frac{1}{2}$), and ($\frac{1}{2}~\frac{1}{2}$~0) present in both but weaker in the needlelike crystal. More details are given in section I of the Supplemental Material (SM). Temperature-dependent measurements confirmed that the modulation was commensurate with the lattice and present over the entire measured temperature range (120-300~K). The additional vectors point to a larger face-centered lattice, but an attempted refinement with such a structure did not improve on that obtained with the expected $Pnma$ unit cell. The nature of this modulation will require further exploration and is beyond the scope of this work; for now we note that WP has the most distorted structure (based on comparison to ``ideal'' orthorhombic lattice parameters) of any MnP-type binary~\cite{CuonoMnPType}. This is a result of the large area associated with the 5$d$ W orbitals~\cite{NigroWP}, leading to more overlap with nearby atoms than in other MnP-type materials, and may have some bearing on the periodic lattice modulation.

\begin{figure*}
    \centering
    \includegraphics[width=0.9\textwidth]{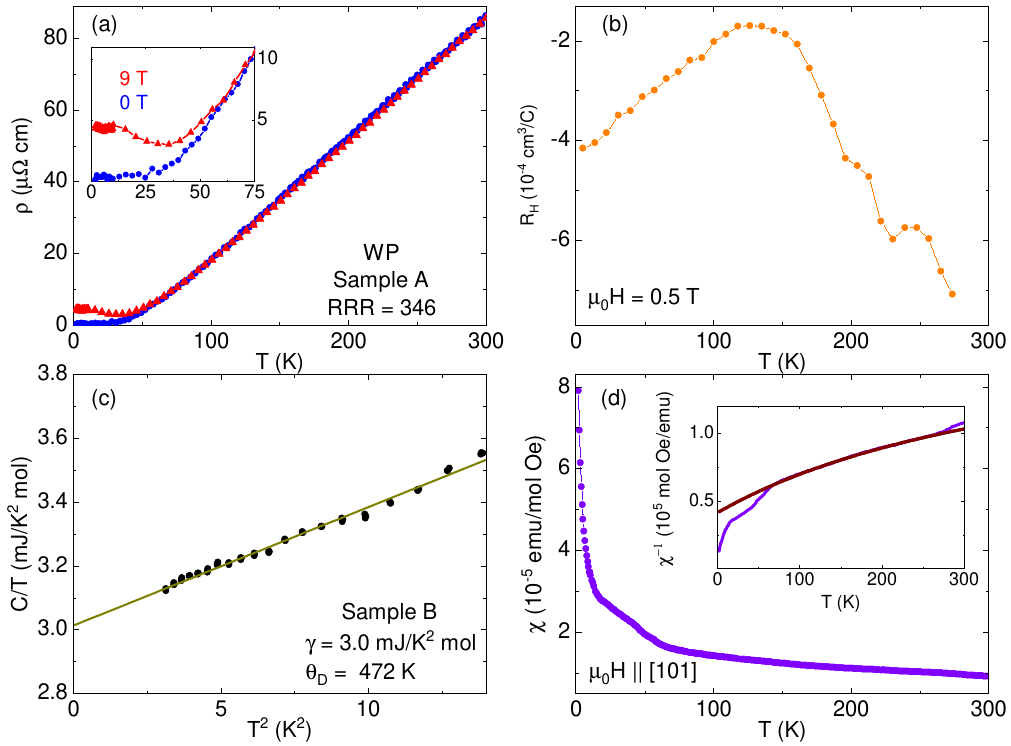}
    \caption{(a) Resistivity as a function of temperature for a WP single crystal with current along the \textit{b}-axis, in zero field and with 9~T applied field. (b) Hall coefficient (based on temperature sweeps in $\pm$14~T) for a WP single crystal. (c) Low temperature specific heat of a different WP single crystal. The green line is a fit to the Debye model, with extracted parameters noted. (d) Magnetization of WP powder as a function of temperature. Data shown are field cooled, but there was no difference with zero field cooling. Inset: an inverse susceptibility plot of the same data with a Curie-Weiss fit (maroon line) over the data from 100-250~K (see text for details). For (a), (b), and (d), field was applied parallel to the [101].}
    \label{fig:WPNormalState}
\end{figure*}

The resistivity has a linear dependence at high temperature, before leveling off below 30~K [Fig.~\ref{fig:WPNormalState}(a)]. This is much like what was previously seen~\cite{LiuWP}, however crystals grown with the WO$_3$ method have substantially lower residual resistivity than those grown from WP polycrystals. The residual resistivity ratio (RRR, defined as $\rho{}_{300~\textrm{K}}/\rho{}_{1.8~\textrm{K}}$) exceeds 300 in some cases, with resistivity $\rho{}$ values down to about 0.2~$\upmu{}\Omega$~cm at base temperature. These are much higher and lower than the initial report on these materials (about 40 and 1~$\upmu{}\Omega{}$~cm, respectively, as the previous work reported a 300~K resistivity that is about half of our value), and we believe they are a direct result of the different growth technique, where the materials are better able to mix in the gas phase. The lowest residual resistivities of other MnP-type phosphides are also about 0.1-0.3~$\upmu{}\Omega{}$~cm~\cite{CampbellFeP,TakaseMnP,NiuCrP}, indicating that our growth technique approaches what may be a rough lower limit of resistivity in this family. Magnetoresistance becomes appreciable around 50~K, and d$\rho$/dT is negative below 35~K in an applied field of 9~T. This is about the same ``turn on'' temperature at which MR becomes significant in isostructural CrP~\cite{NiuCrP} and FeP~\cite{CampbellFeP} as well as topological materials such as WTe$_2$~\cite{AliWTe2}.

In other materials, the minimum seen in $\rho$(\textit{T}) when field is applied is a sign of a change in carrier concentration or mobility ratios of carriers of different sign~\cite{CampbellFeP,AliWTe2,TaftiXMR}. This is often reflected by a complicated temperature dependence to the Hall resistance. However, the large magnetoresistance of these samples at low temperatures makes it difficult to isolate a fully antisymmetric Hall effect signal. Instead, temperature sweeps were made at $\pm$14~T with \textbf{H}~$\parallel$~[101], with the difference at the two field extremes used to calculate a slope, which implicitly assumes a linear Hall resistance [Fig.~\ref{fig:WPNormalState}(b)] but is acceptable for qualitative analysis. Conduction is electron-dominated at all temperatures; band structure calculations anticipate hole carriers as well, though they are expected to be form two dimensional sheets at the Fermi surface~\cite{CuonoMnPType}. There is a clear temperature dependence to the Hall coefficient $R_H$, including a broad maximum at 130~K. There is no clear feature below 50~K, in the vicinity of the increase in MR. Still, the nonmonotonic temperature dependence is similar to what has been seen in other materials with this structure~\cite{CampbellCoAs,CampbellFeP,SegawaFeAs}.

The low temperature specific heat data [Fig.~\ref{fig:WPNormalState}(c)] can be fit reasonably well by the Debye low temperature model $C_p/T = \gamma{} + \beta{}T^2$. The extracted $\gamma$ value of 3.0~$\frac{\textrm{mJ}}{\textrm{mol K}^2}$ is twice that reported for the needlelike samples~\cite{LiuWP}. The Debye temperature calculated from the slope of the $C/T$~vs.~$T^2$ plot is 472~K. The magnetic susceptibility $\chi$ of 94~mg of WP powder taken from the hot end of a growth ampule (to ensure there was no W$_8$P$_4$O$_{32}$) was also measured [Fig.~\ref{fig:WPNormalState}(d)]. Attempts were made with single crystals, but the combination of a small moment and still relatively small mass made it difficult to detect a signal. Data shown are only for field cooling at 0.5~T, but zero field cooling or higher field gave similar results. The susceptibility increases with cooling, with noticeable kinks coming at about 60~K and 15~K. However, there is no indication of long range order. A fit to the data using the Curie-Weiss formula $\chi{} = \chi{}_0 + \frac{C}{T - \Theta{}_{\textrm{CW}}}$ over the range 100-250~K [Fig.~\ref{fig:WPNormalState}(d), inset] gives values of $5.2~\times{}~10^{-6} \frac{\textrm{emu}}{\textrm{mol Oe}}$ for $\chi{}_0$, accounting for parasitic para- and diamagnetic contributions, $1.8~\times{}~10^{-3} \frac{\textrm{emu K}}{\textrm{mol Oe}}$ for the Curie constant $C$, and -92~K for the Curie-Weiss temperature $\Theta_{\textrm{CW}}$, indicating dominant antiferromagnetic fluctuations.

\begin{figure}
    \centering
    \includegraphics[width=0.48\textwidth]{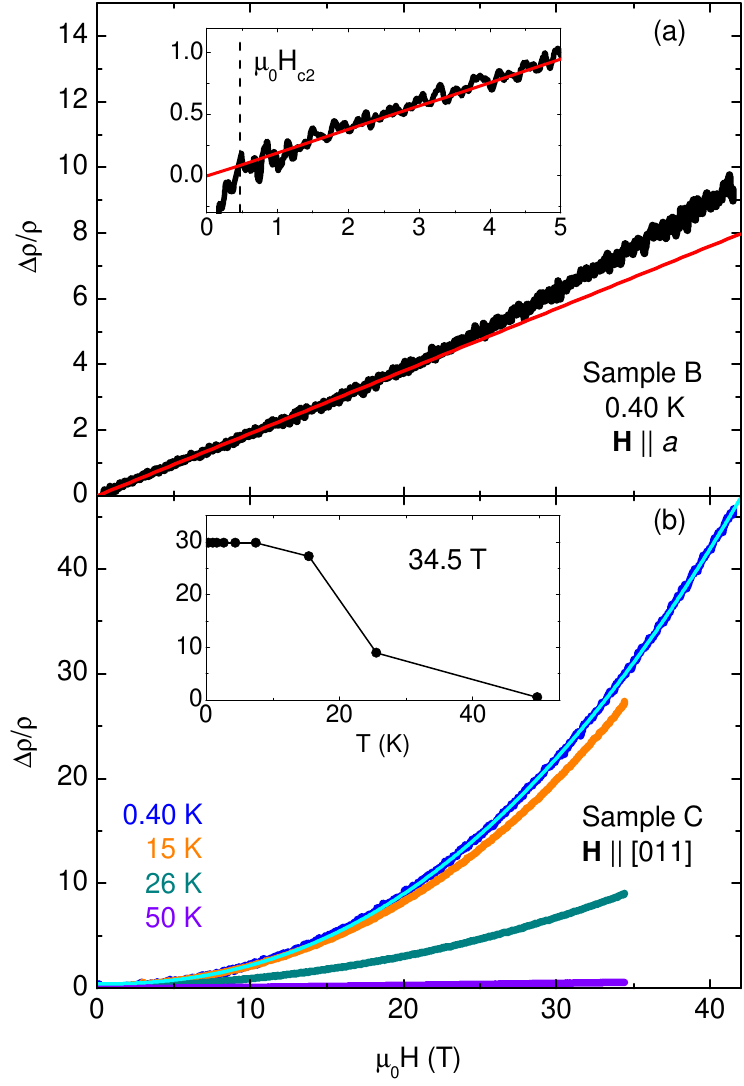}
    \caption{(a) Magnetoresistance (defined as $\frac{\rho(B) - \rho_0}{\rho_0}$, where $\rho_0$ is the lowest resistance above the superconducting transition) of a WP sample with \textbf{H}~$\parallel$~\textit{a}-axis at 0.39~K up to 41.5 T. Data up to 3~K were practically indistinguishable, aside from superconducting transitions visible at very low field in the data below 1.3~K. The red line is a linear fit to normal state data from $H_{c2}$ (determined by the change in slope, see inset) to 20~T. (b) MR of a different WP crystal with \textbf{H}~$\parallel$~[011] at multiple temperatures. Data showed little variation from base temperature to 8~K, and the next lowest temperature was 15.4~K. The light blue line is a power law fit of the base temperature data, yielding \textit{n}~=~2.27. The inset shows MR at 34.5~T (the maximum field of the higher temperature measurements) as a function of temperature, including temperatures not shown in the main plot.}
    \label{fig:WPMagLab}
\end{figure}

Two WP samples were measured at high fields at the NHMFL. One was only measured at very low temperatures with field along the \textit{a}-axis [Fig.~\ref{fig:WPMagLab}(a)]. It showed a nearly linear dependence on field before gradually starting to curve upward above 20~T (a red line marks the 0-20~T fit). Data were nearly identical up to 3~K, the highest measured temperature, with no noticeable decrease in MR at highest field. The increasing width of the signal at high field is due to noise\textemdash{}no coherent quantum oscillations were found. The second sample had its [011] axis aligned with the field, and data were taken over a wider temperature range, allowing the drop in MR with increasing temperature to be seen [Fig.~\ref{fig:WPMagLab}(b)]. A generic power law fit to the lowest temperature data yields \textit{n}~=~2.27, with little deviation over the entire field range. The MR is also about five times larger at base temperature and maximum field for this orientation than for the other angle.

Though we measured two quite different magnetoresistance forms and magnitudes of the MR at high field, we cannot make a comment about MR anisotropy since the data come from separate samples. However, the linearity of Sample B in Fig.~\ref{fig:WPMagLab}(a) up to 20~T is reminiscent of the high field linear magnetoresistance seen in FeP~\cite{CampbellFeP} and (under pressure) CrAs~\cite{NiuCrAs}, also starting from very low field. There it is attributed to ``semi-Dirac points'' in the band structure, points from which the dispersion is linear in a single crystal direction. This linear dispersion in turn can give rise to linear, nonsaturating magnetoresistance when aligned with an applied field. While for those two linearity occurred at \textbf{H}~$\parallel$~\textit{c}, in the absence of magnetic order the same semi-Dirac point will appear at other places in the band structure and be crystallographically protected~\cite{CampbellFeP,CuonoMnPType}. This means that a paramagnetic material such as WP could have linear magnetoresistance in other alignments.


\section{Superconducting Properties}

\begin{figure}
    \centering
    \includegraphics[width=0.48\textwidth]{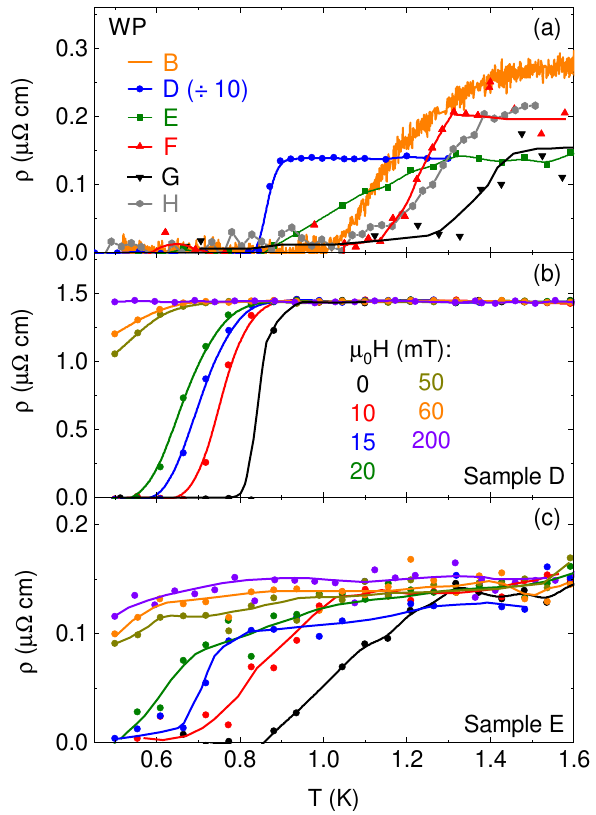}
    \caption{(a) Zero field temperature sweeps, showing superconducting transitions, of five different WP single crystals. Sample D has had its resistivity reduced a factor of 10 to fit the plot scale, while for Sample B, measured at the NHMFL, resistance was not converted to resistivity so the units are arbitrary. The lower two panels are temperature sweeps in various fields for WP samples (b) D and (c) E, which were measured simultaneously. All lines are guides to the eye.}
    \label{fig:WPSC}
\end{figure}

Figure~\ref{fig:WPSC}(a) shows zero-field temperature sweeps for five WP samples, whose superconducting transitions vary in character. Sample D had a higher residual resistivity (1.5~$\upmu\Omega$~cm, reduced by a factor of 10 in the figure) and more needlelike shape, despite being grown with the oxide method. Its superconducting transition at 0.85~K is very sharp, very similar to the first report of superconductivity in this material~\cite{LiuWP}. Four samples with higher RRR values, B and E-G, showed transitions that began and ended above 0.85~K, though they were also much broader. As shwon in Fig.~S2 of the Supplemental Material, this may be a result of Ohmic heating resulting from using high currents to reduce noise. Sample B is the \textbf{H}~$\parallel$~\textit{a}-axis sample from Fig.~\ref{fig:WPMagLab}(a), and the data come from the zero-field cooldown at the NHMFL. The resistance values for that sample were not converted from resistance to resistivity, and so the resistance has simply been scaled to fit onto the plot. The data for samples B and E-H are noisier because the resistance is lower, due to both the lower residual resistivity and the different typical geometric factor of platelike samples. The contrast between the two morphologies can be seen in Fig.~\ref{fig:WPCrystals}(b). While the platelike samples could be polished to try to maximize resistance, this could not be done to the extreme degrees of needlelike samples, which could have an as-grown length over 1~mm and a thickness of less than 50~$\upmu$m. As a result the platelike samples are closer to the noise floor of the measurement systems. This can influence the appearance of the resistive transition, as explored in further detail in Section II of the SM. The elevated $T_c$ is nevertheless apparent. While affected by Ohmic heating from the applied current~[SM, Fig.~S2], the broadness of the transitions seems to be an inherent feature of the samples. There is a clear contrast in transition width with the lower $T_c$ sample (D, Fig~\ref{fig:WPSC}(b)), including for a sample measured at the same time which would have had the same field environment (E, Fig~\ref{fig:WPSC}(c)). In the high field experiment, a superconducting transition was seen in field sweeps at temperatures up to 1.3~K. However, the $\rho{}(B)$ data at very low field for such measurements were not reliable enough for further analysis.

\begin{figure}
    \centering
    \includegraphics[width=0.48\textwidth]{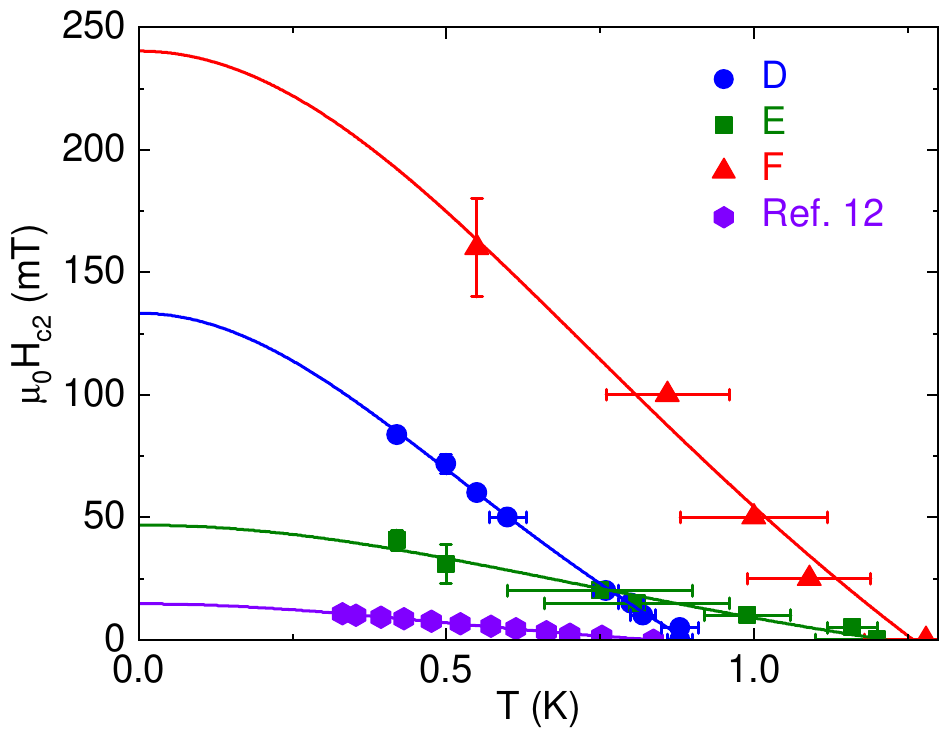}
    \caption{The critical field for the same WP samples shown in Fig.~\ref{fig:WPSC}(a) with matching colors and symbols, as well as the sample from the work of Liu et al.~\cite{LiuWP} with the highest critical field. The criteria for our samples was a 10\% drop in resistivity; for the reference samples the transitions are sharp enough that different criteria would not impact the appearance of the data. Error bars mark uncertainty in either temperature (horizontal) or field (vertical) sweeps during measurements when the other variable was held constant. D and F have \textbf{H}~$\parallel$~\textit{c}-axis, for E and the reference sample the orientations are unknown. Fits are made according to a Ginzburg-Landau formula, $H_{c2}\textrm{(\textit{T})} = H_{c2,0}\Big[\frac{1-(\frac{T}{T_c})^2}{1+(\frac{T}{T_c})^2}\Big]$.}
    \label{fig:WPHc}
\end{figure}

The upper critical field $H_{c2}$ is, like $T_c$, increased in the platelike samples [Fig.~\ref{fig:WPHc}]. Given the broadness of the transitions we took the beginning of the resistance drop (90\% of normal state resistivity) as $H_{c2}$, though some samples still had a slight positive slope in the normal state at low temperature, complicating this analysis [e.g., Fig.~\ref{fig:WPSC}(c)]. Error bars in Fig.~\ref{fig:WPHc} come from an assessment of the noise in the signal. Samples G and H were not measured in field. In comparing samples D and E, measured simultaneously, we see that the broadness of the higher RRR samples is exaggerated by field [Fig.~\ref{fig:WPSC}(b)]: the transition starts at higher temperature for E, but zero resistance occurs at about the same temperature for both. For two samples, D and F, both measured with \textbf{H} along \textit{c}-axis, we see that the higher $T_c$ results in a significantly higher critical field. Even so, in the case of sample D, $H_{c2}$ is roughly five times larger than the highest value reported by Liu et al. for samples with a similar $T_c$ and residual resistivity~\cite{LiuWP}, and in low temperature field sweeps is higher than sample E. This is not just a question of orientation, as rotational measurements on WP crystals grown from polycrystal have shown a change of only about a factor of two with angle~\cite{LiuWP,NigroWP}. The $H_{c2}$ vs. $T_c$ curves fit well (though less so for sample F) to the Ginzburg-Landau formula of the form $H_{c2}\textrm{(\textit{T})} = H_{c2,0}\Big[\frac{1-(\frac{T}{T_c})^2}{1+(\frac{T}{T_c})^2}\Big]$. That being said, our data points are primarily in the higher temperature, more linear portion of the curve.

\section{Discussion}
Variation of $T_c$ for materials grown by CVT has precedent, for example in the case of triplet superconductor UTe$_2$. There, variations in growth conditions can change not only $T_c$ but also the RRR and the appearance of the heat capacity transition~\cite{CairnsUTe2Variation,RosaUTe2Variation,SakaiUTe2Growth}. WP was predicted to be an unconventional topological crystalline superconductor~\cite{CuonoMnPType}. The same is expected for CrAs and (upon suitable doping) MnP, the two isostructural compounds that superconduct under pressure with suppression of magnetic order~\cite{WuCrAs,KotegawaCrAs,ChengMnP}. The competition those two exhibit between magnetic order and superconductivity produces a phase diagram reminiscent of that seen in quantum critical materials~\cite{ChengCrAsMnP}, and in the case of CrAs there is some experimental evidence for triplet superconductivity~\cite{GuoCrAsTriplet,KotegawaCrAsNQR}. A non $s$-wave gap could account for the $T_c$ variation in WP in samples with lower residual resistivity. When the superconducting gap is anisotropic, Cooper pairs can be more easily destroyed by scattering off areas where the gap is very small or zero. Samples with reduced scattering should have higher temperature transitions, as in the case of multicomponent superconductor Sr$_2$RuO$_4$~\cite{MackenzieSr2RuO4,BenhabibSr2RuO4,GhoshSr2RuO4}.

The transitions are broader in the higher RRR samples. As explored further in the Supplemental Material, this may be a result of using higher current values to reduce noise [Fig.~S2]. Though even when lower current was used, tranistions did not appear as sharp as samples with a 0.8~K $T_c$. One possibility is filamentary superconductivity, which could be assessed with more bulk probes. However, while we attempted to measure the superconducting transition in specific heat and with a tunnel diode oscillator technique, in both cases there was no a signal of superconductivity at any temperature, likely due to a low sample mass. Measurements that could confirm bulk superconductivity in oxide-grown crystals are essential for further study. In the absence of that, we have to look elsewhere to determine if the $T_c$ variation is inherent to WP. The XRD measurements found no evidence of any other phases or nonstoichiometry that might produce a spurious, local zero resistance signal. Additionally, the upper critical fields of the higher $T_c$ samples are much larger than those found in the previous study~\cite{LiuWP}. Theoretical calculations for the $T_c$ of WP closely matched the previously observed value of about 0.84~K~\cite{TayranWPTheory}. However, our values of $\gamma{}$ and $\theta{}_D$ are roughly double the previous experiment and theory (about 1.3~mJ/mol K$^2$ and 240~K respectively), which, in the absence of other changes, would raise $T_c$. A Bloch-Gr\"{u}neisen fit of our resistivity data produces a value of the electron-phonon coupling constant $\lambda_{ep}$ that is very close to that obtained in the previous work [SM, Fig.~S3]. However, if superconductivity is indeed unconventional it is questionable how much of a role electron-phonon coupling plays in determining $T_c$.

There is precedent for a higher temperature but broader transition in strongly correlated superconductors. In CeRhIn$_5$, the resistive transition starts well above the temperature determined by heat capacity, and is much broader for current in the $ab$ plane compared to along the \textit{c}-axis~\cite{ParkCeRhIn5TcVariation}, even though the former has an extrapolated normal state residual resistivity about an order of magnitude lower. This is attributed to a change in the antiferromagnetic state slightly above bulk $T_{c}$, and only appears for higher quality samples. Multigap superconductors LaNiC$_2$~\cite{ChenLaNiC2} and SnSb~\cite{LiuSnSb} both have a difference between bulk and resistive $T_c$, with the zero resistance state also having a much higher critical field. SnSb in particular is an interesting case, as it features a superstructure on top of the basic rock salt crystal symmetry, similar to the modulated structure seen in WP. This modulated structure could affect the orbital overlap and electronic configuration slightly so that a transition is visible above the typical superconducting temperature. Band hybridization, spin-orbit coupling, and an anisotropic gap, all expected to be relevant to a proper description of WP~\cite{CuonoMnPType}, could likewise be sensitive to slight variations. While the two samples on which single crystal XRD were performed had nearly identical lattice parameters and both showed the modulation, it was stronger in the three dimensional, higher $T_c$ one.

This last point returns to the connection between growth method, morphology, crystal quality, and superconductivity. Oxide-assisted growth favors more 3D samples than WP powder-based synthesis, and we only find higher temperature resistive transitions in the 3D samples. XRD has shown no difference in the \textit{Pnma} lattice parameters between the two morphologies. The structural differences between the samples are then the modulation strength and the reduced impurity concentration, as indicated by residual resistivity. How directly the two could be linked is unclear. The relation between the possible multiple superconducting phases and the predicted topological crystalline superconductivity is, likewise, something to be explored.

Vapor transport growth using WO$_3$ and elemental P has been found to produce high quality WP single crystals, with low temperature scattering near the lower limit of the MnP family. Oxide-assisted samples also have a higher temperature, but broader, resistive superconducting transition, with a higher critical field. This variation could emerge from the predicted nontrivial topological superconductivity of WP, or the modulated crystal superstructure. Deeper investigation will help sort out the nature of this superconducting phase and its relationship to predicted topological properties as well as the electronic and crystal structure of WP.

\section{\label{sec:Acknowledge}Acknowledgments}

This work was supported by the Department of Energy Award No. DE-SC-0019154 (experimental characterization), the Gordon and Betty Moore Foundation’s EPiQS Initiative through Grant No. GBMF9071 (materials synthesis), and the Maryland Quantum Materials Center. D.J.C. was supported in part by the Anne G. Wylie Dissertation Fellowship. A portion of this work was performed at the National High Magnetic Field Laboratory, which is supported by National Science Foundation Cooperative Agreement no.~DMR-1644779 as well as the state of Florida. We also acknowledge the support of the Maryland NanoCenter and its FabLab. This work was performed under the auspices of the U.S. Department of Energy by Lawrence Livermore National Laboratory under Contract DE-AC52-07NA27344.

\bibliography{WP}

\begin{thebibliography}{42}%
\makeatletter
\providecommand \@ifxundefined [1]{%
 \@ifx{#1\undefined}
}%
\providecommand \@ifnum [1]{%
 \ifnum #1\expandafter \@firstoftwo
 \else \expandafter \@secondoftwo
 \fi
}%
\providecommand \@ifx [1]{%
 \ifx #1\expandafter \@firstoftwo
 \else \expandafter \@secondoftwo
 \fi
}%
\providecommand \natexlab [1]{#1}%
\providecommand \enquote  [1]{``#1''}%
\providecommand \bibnamefont  [1]{#1}%
\providecommand \bibfnamefont [1]{#1}%
\providecommand \citenamefont [1]{#1}%
\providecommand \href@noop [0]{\@secondoftwo}%
\providecommand \href [0]{\begingroup \@sanitize@url \@href}%
\providecommand \@href[1]{\@@startlink{#1}\@@href}%
\providecommand \@@href[1]{\endgroup#1\@@endlink}%
\providecommand \@sanitize@url [0]{\catcode `\\12\catcode `\$12\catcode
  `\&12\catcode `\#12\catcode `\^12\catcode `\_12\catcode `\%12\relax}%
\providecommand \@@startlink[1]{}%
\providecommand \@@endlink[0]{}%
\providecommand \url  [0]{\begingroup\@sanitize@url \@url }%
\providecommand \@url [1]{\endgroup\@href {#1}{\urlprefix }}%
\providecommand \urlprefix  [0]{URL }%
\providecommand \Eprint [0]{\href }%
\providecommand \doibase [0]{http://dx.doi.org/}%
\providecommand \selectlanguage [0]{\@gobble}%
\providecommand \bibinfo  [0]{\@secondoftwo}%
\providecommand \bibfield  [0]{\@secondoftwo}%
\providecommand \translation [1]{[#1]}%
\providecommand \BibitemOpen [0]{}%
\providecommand \bibitemStop [0]{}%
\providecommand \bibitemNoStop [0]{.\EOS\space}%
\providecommand \EOS [0]{\spacefactor3000\relax}%
\providecommand \BibitemShut  [1]{\csname bibitem#1\endcsname}%
\let\auto@bib@innerbib\@empty
\bibitem [{\citenamefont {Rundqvist}(1962)}]{RundqvistPhosphides}%
  \BibitemOpen
  \bibfield  {author} {\bibinfo {author} {\bibfnamefont {S.}~\bibnamefont
  {Rundqvist}},\ }\href {\doibase 10.3891/acta.chem.scand.16-0287} {\bibfield
  {journal} {\bibinfo  {journal} {Acta Chem. Scand.}\ }\textbf {\bibinfo
  {volume} {16}},\ \bibinfo {pages} {287} (\bibinfo {year} {1962})}\BibitemShut
  {NoStop}%
\bibitem [{\citenamefont {Rundqvist}\ and\ \citenamefont
  {Lundstrom}(1963)}]{RundqvistWP}%
  \BibitemOpen
  \bibfield  {author} {\bibinfo {author} {\bibfnamefont {S.}~\bibnamefont
  {Rundqvist}}\ and\ \bibinfo {author} {\bibfnamefont {T.}~\bibnamefont
  {Lundstrom}},\ }\href {\doibase 10.3891/acta.chem.scand.17-0037} {\bibfield
  {journal} {\bibinfo  {journal} {Acta Chem. Scand.}\ }\textbf {\bibinfo
  {volume} {17}},\ \bibinfo {pages} {37} (\bibinfo {year} {1963})}\BibitemShut
  {NoStop}%
\bibitem [{\citenamefont {Bellavance}\ \emph {et~al.}(1969)\citenamefont
  {Bellavance}, \citenamefont {Vlasse}, \citenamefont {Morris},\ and\
  \citenamefont {Wold}}]{BellavanceFePPreparation}%
  \BibitemOpen
  \bibfield  {author} {\bibinfo {author} {\bibfnamefont {D.}~\bibnamefont
  {Bellavance}}, \bibinfo {author} {\bibfnamefont {M.}~\bibnamefont {Vlasse}},
  \bibinfo {author} {\bibfnamefont {B.}~\bibnamefont {Morris}}, \ and\ \bibinfo
  {author} {\bibfnamefont {A.}~\bibnamefont {Wold}},\ }\href {\doibase
  https://doi.org/10.1016/0022-4596(69)90011-5} {\bibfield  {journal} {\bibinfo
   {journal} {J. Solid State Chem.}\ }\textbf {\bibinfo {volume} {1}},\
  \bibinfo {pages} {82 } (\bibinfo {year} {1969})}\BibitemShut {NoStop}%
\bibitem [{\citenamefont {Rodriguez}\ \emph {et~al.}(2011)\citenamefont
  {Rodriguez}, \citenamefont {Stock}, \citenamefont {Krycka}, \citenamefont
  {Majkrzak}, \citenamefont {Zajdel}, \citenamefont {Kirshenbaum},
  \citenamefont {Butch}, \citenamefont {Saha}, \citenamefont {Paglione},\ and\
  \citenamefont {Green}}]{RodriguezFeAsSDW}%
  \BibitemOpen
  \bibfield  {author} {\bibinfo {author} {\bibfnamefont {E.~E.}\ \bibnamefont
  {Rodriguez}}, \bibinfo {author} {\bibfnamefont {C.}~\bibnamefont {Stock}},
  \bibinfo {author} {\bibfnamefont {K.~L.}\ \bibnamefont {Krycka}}, \bibinfo
  {author} {\bibfnamefont {C.~F.}\ \bibnamefont {Majkrzak}}, \bibinfo {author}
  {\bibfnamefont {P.}~\bibnamefont {Zajdel}}, \bibinfo {author} {\bibfnamefont
  {K.}~\bibnamefont {Kirshenbaum}}, \bibinfo {author} {\bibfnamefont {N.~P.}\
  \bibnamefont {Butch}}, \bibinfo {author} {\bibfnamefont {S.~R.}\ \bibnamefont
  {Saha}}, \bibinfo {author} {\bibfnamefont {J.}~\bibnamefont {Paglione}}, \
  and\ \bibinfo {author} {\bibfnamefont {M.~A.}\ \bibnamefont {Green}},\ }\href
  {\doibase 10.1103/PhysRevB.83.134438} {\bibfield  {journal} {\bibinfo
  {journal} {Phys. Rev. B}\ }\textbf {\bibinfo {volume} {83}},\ \bibinfo
  {pages} {134438} (\bibinfo {year} {2011})}\BibitemShut {NoStop}%
\bibitem [{\citenamefont {Hirai}\ \emph {et~al.}(2012)\citenamefont {Hirai},
  \citenamefont {Takayama}, \citenamefont {Hashizume},\ and\ \citenamefont
  {Takagi}}]{HiraiRuRhPn}%
  \BibitemOpen
  \bibfield  {author} {\bibinfo {author} {\bibfnamefont {D.}~\bibnamefont
  {Hirai}}, \bibinfo {author} {\bibfnamefont {T.}~\bibnamefont {Takayama}},
  \bibinfo {author} {\bibfnamefont {D.}~\bibnamefont {Hashizume}}, \ and\
  \bibinfo {author} {\bibfnamefont {H.}~\bibnamefont {Takagi}},\ }\href
  {\doibase 10.1103/PhysRevB.85.140509} {\bibfield  {journal} {\bibinfo
  {journal} {Phys. Rev. B}\ }\textbf {\bibinfo {volume} {85}},\ \bibinfo
  {pages} {140509(R)} (\bibinfo {year} {2012})}\BibitemShut {NoStop}%
\bibitem [{\citenamefont {Wu}\ \emph {et~al.}(2014)\citenamefont {Wu},
  \citenamefont {Cheng}, \citenamefont {Matsubayashi}, \citenamefont {Kong},
  \citenamefont {Lin}, \citenamefont {Jin}, \citenamefont {Wang}, \citenamefont
  {Uwatoko},\ and\ \citenamefont {Luo}}]{WuCrAs}%
  \BibitemOpen
  \bibfield  {author} {\bibinfo {author} {\bibfnamefont {W.}~\bibnamefont
  {Wu}}, \bibinfo {author} {\bibfnamefont {J.}~\bibnamefont {Cheng}}, \bibinfo
  {author} {\bibfnamefont {K.}~\bibnamefont {Matsubayashi}}, \bibinfo {author}
  {\bibfnamefont {P.}~\bibnamefont {Kong}}, \bibinfo {author} {\bibfnamefont
  {F.}~\bibnamefont {Lin}}, \bibinfo {author} {\bibfnamefont {C.}~\bibnamefont
  {Jin}}, \bibinfo {author} {\bibfnamefont {N.}~\bibnamefont {Wang}}, \bibinfo
  {author} {\bibfnamefont {Y.}~\bibnamefont {Uwatoko}}, \ and\ \bibinfo
  {author} {\bibfnamefont {J.}~\bibnamefont {Luo}},\ }\href {\doibase
  10.1038/ncomms6508} {\bibfield  {journal} {\bibinfo  {journal} {Nat.
  Commun.}\ }\textbf {\bibinfo {volume} {5}},\ \bibinfo {pages} {5508}
  (\bibinfo {year} {2014})}\BibitemShut {NoStop}%
\bibitem [{\citenamefont {Kotegawa}\ \emph {et~al.}(2014)\citenamefont
  {Kotegawa}, \citenamefont {Nakahara}, \citenamefont {Tou},\ and\
  \citenamefont {Sugawara}}]{KotegawaCrAs}%
  \BibitemOpen
  \bibfield  {author} {\bibinfo {author} {\bibfnamefont {H.}~\bibnamefont
  {Kotegawa}}, \bibinfo {author} {\bibfnamefont {S.}~\bibnamefont {Nakahara}},
  \bibinfo {author} {\bibfnamefont {H.}~\bibnamefont {Tou}}, \ and\ \bibinfo
  {author} {\bibfnamefont {H.}~\bibnamefont {Sugawara}},\ }\href
  {https://doi.org/10.7566/JPSJ.83.093702} {\bibfield  {journal} {\bibinfo
  {journal} {J. Phys. Soc. Jpn.}\ }\textbf {\bibinfo {volume} {83}},\ \bibinfo
  {pages} {093702} (\bibinfo {year} {2014})}\BibitemShut {NoStop}%
\bibitem [{\citenamefont {Cheng}\ \emph {et~al.}(2015)\citenamefont {Cheng},
  \citenamefont {Matsubayashi}, \citenamefont {Wu}, \citenamefont {Sun},
  \citenamefont {Lin}, \citenamefont {Luo},\ and\ \citenamefont
  {Uwatoko}}]{ChengMnP}%
  \BibitemOpen
  \bibfield  {author} {\bibinfo {author} {\bibfnamefont {J.-G.}\ \bibnamefont
  {Cheng}}, \bibinfo {author} {\bibfnamefont {K.}~\bibnamefont {Matsubayashi}},
  \bibinfo {author} {\bibfnamefont {W.}~\bibnamefont {Wu}}, \bibinfo {author}
  {\bibfnamefont {J.~P.}\ \bibnamefont {Sun}}, \bibinfo {author} {\bibfnamefont
  {F.~K.}\ \bibnamefont {Lin}}, \bibinfo {author} {\bibfnamefont {J.~L.}\
  \bibnamefont {Luo}}, \ and\ \bibinfo {author} {\bibfnamefont
  {Y.}~\bibnamefont {Uwatoko}},\ }\href {\doibase
  10.1103/PhysRevLett.114.117001} {\bibfield  {journal} {\bibinfo  {journal}
  {Phys. Rev. Lett.}\ }\textbf {\bibinfo {volume} {114}},\ \bibinfo {pages}
  {117001} (\bibinfo {year} {2015})}\BibitemShut {NoStop}%
\bibitem [{\citenamefont {Niu}\ \emph {et~al.}(2017)\citenamefont {Niu},
  \citenamefont {Yu}, \citenamefont {Yip}, \citenamefont {Lim}, \citenamefont
  {Kotegawa}, \citenamefont {Matsuoka}, \citenamefont {Sugawara}, \citenamefont
  {Tou}, \citenamefont {Yanase},\ and\ \citenamefont {Goh}}]{NiuCrAs}%
  \BibitemOpen
  \bibfield  {author} {\bibinfo {author} {\bibfnamefont {Q.}~\bibnamefont
  {Niu}}, \bibinfo {author} {\bibfnamefont {W.}~\bibnamefont {Yu}}, \bibinfo
  {author} {\bibfnamefont {K.}~\bibnamefont {Yip}}, \bibinfo {author}
  {\bibfnamefont {Z.}~\bibnamefont {Lim}}, \bibinfo {author} {\bibfnamefont
  {H.}~\bibnamefont {Kotegawa}}, \bibinfo {author} {\bibfnamefont
  {E.}~\bibnamefont {Matsuoka}}, \bibinfo {author} {\bibfnamefont
  {H.}~\bibnamefont {Sugawara}}, \bibinfo {author} {\bibfnamefont
  {H.}~\bibnamefont {Tou}}, \bibinfo {author} {\bibfnamefont {Y.}~\bibnamefont
  {Yanase}}, \ and\ \bibinfo {author} {\bibfnamefont {S.~K.}\ \bibnamefont
  {Goh}},\ }\href {\doibase https://doi.org/10.1038/ncomms15358} {\bibfield
  {journal} {\bibinfo  {journal} {Nat. Commun.}\ }\textbf {\bibinfo {volume}
  {8}},\ \bibinfo {pages} {15358} (\bibinfo {year} {2017})}\BibitemShut
  {NoStop}%
\bibitem [{\citenamefont {Cuono}\ \emph {et~al.}(2019)\citenamefont {Cuono},
  \citenamefont {Forte}, \citenamefont {Cuoco}, \citenamefont {Islam},
  \citenamefont {Luo}, \citenamefont {Noce},\ and\ \citenamefont
  {Autieri}}]{CuonoMnPType}%
  \BibitemOpen
  \bibfield  {author} {\bibinfo {author} {\bibfnamefont {G.}~\bibnamefont
  {Cuono}}, \bibinfo {author} {\bibfnamefont {F.}~\bibnamefont {Forte}},
  \bibinfo {author} {\bibfnamefont {M.}~\bibnamefont {Cuoco}}, \bibinfo
  {author} {\bibfnamefont {R.}~\bibnamefont {Islam}}, \bibinfo {author}
  {\bibfnamefont {J.}~\bibnamefont {Luo}}, \bibinfo {author} {\bibfnamefont
  {C.}~\bibnamefont {Noce}}, \ and\ \bibinfo {author} {\bibfnamefont
  {C.}~\bibnamefont {Autieri}},\ }\href {\doibase
  10.1103/PhysRevMaterials.3.095004} {\bibfield  {journal} {\bibinfo  {journal}
  {Phys. Rev. Mater.}\ }\textbf {\bibinfo {volume} {3}},\ \bibinfo {pages}
  {095004} (\bibinfo {year} {2019})}\BibitemShut {NoStop}%
\bibitem [{\citenamefont {Campbell}\ \emph {et~al.}(2021)\citenamefont
  {Campbell}, \citenamefont {Collini}, \citenamefont {S{\l}awi{\'n}ska},
  \citenamefont {Autieri}, \citenamefont {Wang}, \citenamefont {Wang},
  \citenamefont {Wilfong}, \citenamefont {Eo}, \citenamefont {Neves},
  \citenamefont {Graf} \emph {et~al.}}]{CampbellFeP}%
  \BibitemOpen
  \bibfield  {author} {\bibinfo {author} {\bibfnamefont {D.}~\bibnamefont
  {Campbell}}, \bibinfo {author} {\bibfnamefont {J.}~\bibnamefont {Collini}},
  \bibinfo {author} {\bibfnamefont {J.}~\bibnamefont {S{\l}awi{\'n}ska}},
  \bibinfo {author} {\bibfnamefont {C.}~\bibnamefont {Autieri}}, \bibinfo
  {author} {\bibfnamefont {L.}~\bibnamefont {Wang}}, \bibinfo {author}
  {\bibfnamefont {K.}~\bibnamefont {Wang}}, \bibinfo {author} {\bibfnamefont
  {B.}~\bibnamefont {Wilfong}}, \bibinfo {author} {\bibfnamefont
  {Y.}~\bibnamefont {Eo}}, \bibinfo {author} {\bibfnamefont {P.}~\bibnamefont
  {Neves}}, \bibinfo {author} {\bibfnamefont {D.}~\bibnamefont {Graf}},  \emph
  {et~al.},\ }\href {https://doi.org/10.1038/s41535-021-00337-2} {\bibfield
  {journal} {\bibinfo  {journal} {npj Quantum Mater.}\ }\textbf {\bibinfo
  {volume} {6}},\ \bibinfo {pages} {38} (\bibinfo {year} {2021})}\BibitemShut
  {NoStop}%
\bibitem [{\citenamefont {Liu}\ \emph {et~al.}(2019)\citenamefont {Liu},
  \citenamefont {Wu}, \citenamefont {Zhao}, \citenamefont {Zhao}, \citenamefont
  {Cui}, \citenamefont {Shan}, \citenamefont {Zhang}, \citenamefont {Yang},
  \citenamefont {Sun}, \citenamefont {Wei}, \citenamefont {Li}, \citenamefont
  {Zhao}, \citenamefont {Sui}, \citenamefont {Cheng}, \citenamefont {Lu},
  \citenamefont {Luo},\ and\ \citenamefont {Liu}}]{LiuWP}%
  \BibitemOpen
  \bibfield  {author} {\bibinfo {author} {\bibfnamefont {Z.}~\bibnamefont
  {Liu}}, \bibinfo {author} {\bibfnamefont {W.}~\bibnamefont {Wu}}, \bibinfo
  {author} {\bibfnamefont {Z.}~\bibnamefont {Zhao}}, \bibinfo {author}
  {\bibfnamefont {H.}~\bibnamefont {Zhao}}, \bibinfo {author} {\bibfnamefont
  {J.}~\bibnamefont {Cui}}, \bibinfo {author} {\bibfnamefont {P.}~\bibnamefont
  {Shan}}, \bibinfo {author} {\bibfnamefont {J.}~\bibnamefont {Zhang}},
  \bibinfo {author} {\bibfnamefont {C.}~\bibnamefont {Yang}}, \bibinfo {author}
  {\bibfnamefont {P.}~\bibnamefont {Sun}}, \bibinfo {author} {\bibfnamefont
  {Y.}~\bibnamefont {Wei}}, \bibinfo {author} {\bibfnamefont {S.}~\bibnamefont
  {Li}}, \bibinfo {author} {\bibfnamefont {J.}~\bibnamefont {Zhao}}, \bibinfo
  {author} {\bibfnamefont {Y.}~\bibnamefont {Sui}}, \bibinfo {author}
  {\bibfnamefont {J.}~\bibnamefont {Cheng}}, \bibinfo {author} {\bibfnamefont
  {L.}~\bibnamefont {Lu}}, \bibinfo {author} {\bibfnamefont {J.}~\bibnamefont
  {Luo}}, \ and\ \bibinfo {author} {\bibfnamefont {G.}~\bibnamefont {Liu}},\
  }\href {\doibase 10.1103/PhysRevB.99.184509} {\bibfield  {journal} {\bibinfo
  {journal} {Phys. Rev. B}\ }\textbf {\bibinfo {volume} {99}},\ \bibinfo
  {pages} {184509} (\bibinfo {year} {2019})}\BibitemShut {NoStop}%
\bibitem [{\citenamefont {Nigro}\ \emph {et~al.}(2022)\citenamefont {Nigro},
  \citenamefont {Cuono}, \citenamefont {Marra}, \citenamefont {Leo},
  \citenamefont {Grimaldi}, \citenamefont {Liu}, \citenamefont {Mi},
  \citenamefont {Wu}, \citenamefont {Liu}, \citenamefont {Autieri},
  \citenamefont {Luo},\ and\ \citenamefont {Noce}}]{NigroWP}%
  \BibitemOpen
  \bibfield  {author} {\bibinfo {author} {\bibfnamefont {A.}~\bibnamefont
  {Nigro}}, \bibinfo {author} {\bibfnamefont {G.}~\bibnamefont {Cuono}},
  \bibinfo {author} {\bibfnamefont {P.}~\bibnamefont {Marra}}, \bibinfo
  {author} {\bibfnamefont {A.}~\bibnamefont {Leo}}, \bibinfo {author}
  {\bibfnamefont {G.}~\bibnamefont {Grimaldi}}, \bibinfo {author}
  {\bibfnamefont {Z.}~\bibnamefont {Liu}}, \bibinfo {author} {\bibfnamefont
  {Z.}~\bibnamefont {Mi}}, \bibinfo {author} {\bibfnamefont {W.}~\bibnamefont
  {Wu}}, \bibinfo {author} {\bibfnamefont {G.}~\bibnamefont {Liu}}, \bibinfo
  {author} {\bibfnamefont {C.}~\bibnamefont {Autieri}}, \bibinfo {author}
  {\bibfnamefont {J.}~\bibnamefont {Luo}}, \ and\ \bibinfo {author}
  {\bibfnamefont {C.}~\bibnamefont {Noce}},\ }\href
  {https://www.mdpi.com/1996-1944/15/3/1027} {\bibfield  {journal} {\bibinfo
  {journal} {Materials}\ }\textbf {\bibinfo {volume} {15}} (\bibinfo {year}
  {2022})}\BibitemShut {NoStop}%
\bibitem [{\citenamefont {Mackenzie}\ \emph {et~al.}(1998)\citenamefont
  {Mackenzie}, \citenamefont {Haselwimmer}, \citenamefont {Tyler},
  \citenamefont {Lonzarich}, \citenamefont {Mori}, \citenamefont {Nishizaki},\
  and\ \citenamefont {Maeno}}]{MackenzieSr2RuO4}%
  \BibitemOpen
  \bibfield  {author} {\bibinfo {author} {\bibfnamefont {A.~P.}\ \bibnamefont
  {Mackenzie}}, \bibinfo {author} {\bibfnamefont {R.~K.~W.}\ \bibnamefont
  {Haselwimmer}}, \bibinfo {author} {\bibfnamefont {A.~W.}\ \bibnamefont
  {Tyler}}, \bibinfo {author} {\bibfnamefont {G.~G.}\ \bibnamefont
  {Lonzarich}}, \bibinfo {author} {\bibfnamefont {Y.}~\bibnamefont {Mori}},
  \bibinfo {author} {\bibfnamefont {S.}~\bibnamefont {Nishizaki}}, \ and\
  \bibinfo {author} {\bibfnamefont {Y.}~\bibnamefont {Maeno}},\ }\href
  {\doibase 10.1103/PhysRevLett.80.161} {\bibfield  {journal} {\bibinfo
  {journal} {Phys. Rev. Lett.}\ }\textbf {\bibinfo {volume} {80}},\ \bibinfo
  {pages} {161} (\bibinfo {year} {1998})}\BibitemShut {NoStop}%
\bibitem [{\citenamefont {Park}\ \emph {et~al.}(2012)\citenamefont {Park},
  \citenamefont {Lee}, \citenamefont {Martin}, \citenamefont {Lu},
  \citenamefont {Sidorov}, \citenamefont {Gofryk}, \citenamefont {Ronning},
  \citenamefont {Bauer},\ and\ \citenamefont
  {Thompson}}]{ParkCeRhIn5TcVariation}%
  \BibitemOpen
  \bibfield  {author} {\bibinfo {author} {\bibfnamefont {T.}~\bibnamefont
  {Park}}, \bibinfo {author} {\bibfnamefont {H.}~\bibnamefont {Lee}}, \bibinfo
  {author} {\bibfnamefont {I.}~\bibnamefont {Martin}}, \bibinfo {author}
  {\bibfnamefont {X.}~\bibnamefont {Lu}}, \bibinfo {author} {\bibfnamefont
  {V.~A.}\ \bibnamefont {Sidorov}}, \bibinfo {author} {\bibfnamefont
  {K.}~\bibnamefont {Gofryk}}, \bibinfo {author} {\bibfnamefont
  {F.}~\bibnamefont {Ronning}}, \bibinfo {author} {\bibfnamefont {E.~D.}\
  \bibnamefont {Bauer}}, \ and\ \bibinfo {author} {\bibfnamefont {J.~D.}\
  \bibnamefont {Thompson}},\ }\href {\doibase 10.1103/PhysRevLett.108.077003}
  {\bibfield  {journal} {\bibinfo  {journal} {Phys. Rev. Lett.}\ }\textbf
  {\bibinfo {volume} {108}},\ \bibinfo {pages} {077003} (\bibinfo {year}
  {2012})}\BibitemShut {NoStop}%
\bibitem [{\citenamefont {Chen}\ \emph {et~al.}(2013)\citenamefont {Chen},
  \citenamefont {Jiao}, \citenamefont {Zhang}, \citenamefont {Chen},
  \citenamefont {Yang}, \citenamefont {Nicklas}, \citenamefont {Steglich},\
  and\ \citenamefont {Yuan}}]{ChenLaNiC2}%
  \BibitemOpen
  \bibfield  {author} {\bibinfo {author} {\bibfnamefont {J.}~\bibnamefont
  {Chen}}, \bibinfo {author} {\bibfnamefont {L.}~\bibnamefont {Jiao}}, \bibinfo
  {author} {\bibfnamefont {J.~L.}\ \bibnamefont {Zhang}}, \bibinfo {author}
  {\bibfnamefont {Y.}~\bibnamefont {Chen}}, \bibinfo {author} {\bibfnamefont
  {L.}~\bibnamefont {Yang}}, \bibinfo {author} {\bibfnamefont {M.}~\bibnamefont
  {Nicklas}}, \bibinfo {author} {\bibfnamefont {F.}~\bibnamefont {Steglich}}, \
  and\ \bibinfo {author} {\bibfnamefont {H.~Q.}\ \bibnamefont {Yuan}},\ }\href
  {\doibase 10.1088/1367-2630/15/5/053005} {\bibfield  {journal} {\bibinfo
  {journal} {New J. Phys.}\ }\textbf {\bibinfo {volume} {15}},\ \bibinfo
  {pages} {053005} (\bibinfo {year} {2013})}\BibitemShut {NoStop}%
\bibitem [{\citenamefont {Liu}\ \emph {et~al.}(2018)\citenamefont {Liu},
  \citenamefont {Wu}, \citenamefont {Cui}, \citenamefont {Wang}, \citenamefont
  {Liu}, \citenamefont {Wang}, \citenamefont {Ren},\ and\ \citenamefont
  {Cao}}]{LiuSnSb}%
  \BibitemOpen
  \bibfield  {author} {\bibinfo {author} {\bibfnamefont {B.}~\bibnamefont
  {Liu}}, \bibinfo {author} {\bibfnamefont {J.}~\bibnamefont {Wu}}, \bibinfo
  {author} {\bibfnamefont {Y.}~\bibnamefont {Cui}}, \bibinfo {author}
  {\bibfnamefont {H.}~\bibnamefont {Wang}}, \bibinfo {author} {\bibfnamefont
  {Y.}~\bibnamefont {Liu}}, \bibinfo {author} {\bibfnamefont {Z.}~\bibnamefont
  {Wang}}, \bibinfo {author} {\bibfnamefont {Z.}~\bibnamefont {Ren}}, \ and\
  \bibinfo {author} {\bibfnamefont {G.}~\bibnamefont {Cao}},\ }\href {\doibase
  10.1088/1361-6668/aae6fe} {\bibfield  {journal} {\bibinfo  {journal}
  {Supercond. Sci. Technol.}\ }\textbf {\bibinfo {volume} {31}},\ \bibinfo
  {pages} {125011} (\bibinfo {year} {2018})}\BibitemShut {NoStop}%
\bibitem [{\citenamefont {Krause}\ \emph {et~al.}(2015)\citenamefont {Krause},
  \citenamefont {Herbst-Irmer}, \citenamefont {Sheldrick},\ and\ \citenamefont
  {Stalke}}]{KrauseStructure}%
  \BibitemOpen
  \bibfield  {author} {\bibinfo {author} {\bibfnamefont {L.}~\bibnamefont
  {Krause}}, \bibinfo {author} {\bibfnamefont {R.}~\bibnamefont
  {Herbst-Irmer}}, \bibinfo {author} {\bibfnamefont {G.~M.}\ \bibnamefont
  {Sheldrick}}, \ and\ \bibinfo {author} {\bibfnamefont {D.}~\bibnamefont
  {Stalke}},\ }\href {\doibase 10.1107/S1600576714022985} {\bibfield  {journal}
  {\bibinfo  {journal} {J. Appl. Crystallogr.}\ }\textbf {\bibinfo {volume}
  {48}},\ \bibinfo {pages} {3} (\bibinfo {year} {2015})}\BibitemShut {NoStop}%
\bibitem [{\citenamefont {Sheldrick}(2015{\natexlab{a}})}]{SheldrickShelXT}%
  \BibitemOpen
  \bibfield  {author} {\bibinfo {author} {\bibfnamefont {G.~M.}\ \bibnamefont
  {Sheldrick}},\ }\href@noop {} {\bibfield  {journal} {\bibinfo  {journal}
  {Acta Crystallogr. A}\ }\textbf {\bibinfo {volume} {71}},\ \bibinfo {pages}
  {3} (\bibinfo {year} {2015}{\natexlab{a}})}\BibitemShut {NoStop}%
\bibitem [{\citenamefont {Sheldrick}(2015{\natexlab{b}})}]{SheldrickShelXL}%
  \BibitemOpen
  \bibfield  {author} {\bibinfo {author} {\bibfnamefont {G.~M.}\ \bibnamefont
  {Sheldrick}},\ }\href@noop {} {\bibfield  {journal} {\bibinfo  {journal}
  {Acta Crystallogr. C}\ }\textbf {\bibinfo {volume} {71}},\ \bibinfo {pages}
  {3} (\bibinfo {year} {2015}{\natexlab{b}})}\BibitemShut {NoStop}%
\bibitem [{\citenamefont {Binnewies}\ \emph {et~al.}(2011)\citenamefont
  {Binnewies}, \citenamefont {Glaum}, \citenamefont {Schmidt},\ and\
  \citenamefont {Schmidt}}]{BinnewiesCVTBook}%
  \BibitemOpen
  \bibfield  {author} {\bibinfo {author} {\bibfnamefont {M.}~\bibnamefont
  {Binnewies}}, \bibinfo {author} {\bibfnamefont {R.}~\bibnamefont {Glaum}},
  \bibinfo {author} {\bibfnamefont {M.}~\bibnamefont {Schmidt}}, \ and\
  \bibinfo {author} {\bibfnamefont {P.}~\bibnamefont {Schmidt}},\ }\href@noop
  {} {\emph {\bibinfo {title} {{Chemische Transportreaktionen}}}}\ (\bibinfo
  {publisher} {Walter de Gruyter},\ \bibinfo {address} {Berlin, Germany},\
  \bibinfo {year} {2011})\BibitemShut {NoStop}%
\bibitem [{\citenamefont {Niu}\ \emph {et~al.}(2019)\citenamefont {Niu},
  \citenamefont {Yu}, \citenamefont {Aulestia}, \citenamefont {Hu},
  \citenamefont {Lai}, \citenamefont {Kotegawa}, \citenamefont {Matsuoka},
  \citenamefont {Sugawara}, \citenamefont {Tou}, \citenamefont {Sun},
  \citenamefont {Balakirev}, \citenamefont {Yanase},\ and\ \citenamefont
  {Goh}}]{NiuCrP}%
  \BibitemOpen
  \bibfield  {author} {\bibinfo {author} {\bibfnamefont {Q.}~\bibnamefont
  {Niu}}, \bibinfo {author} {\bibfnamefont {W.~C.}\ \bibnamefont {Yu}},
  \bibinfo {author} {\bibfnamefont {E.~I.~P.}\ \bibnamefont {Aulestia}},
  \bibinfo {author} {\bibfnamefont {Y.~J.}\ \bibnamefont {Hu}}, \bibinfo
  {author} {\bibfnamefont {K.~T.}\ \bibnamefont {Lai}}, \bibinfo {author}
  {\bibfnamefont {H.}~\bibnamefont {Kotegawa}}, \bibinfo {author}
  {\bibfnamefont {E.}~\bibnamefont {Matsuoka}}, \bibinfo {author}
  {\bibfnamefont {H.}~\bibnamefont {Sugawara}}, \bibinfo {author}
  {\bibfnamefont {H.}~\bibnamefont {Tou}}, \bibinfo {author} {\bibfnamefont
  {D.}~\bibnamefont {Sun}}, \bibinfo {author} {\bibfnamefont {F.~F.}\
  \bibnamefont {Balakirev}}, \bibinfo {author} {\bibfnamefont {Y.}~\bibnamefont
  {Yanase}}, \ and\ \bibinfo {author} {\bibfnamefont {S.~K.}\ \bibnamefont
  {Goh}},\ }\href {\doibase 10.1103/PhysRevB.99.125126} {\bibfield  {journal}
  {\bibinfo  {journal} {Phys. Rev. B}\ }\textbf {\bibinfo {volume} {99}},\
  \bibinfo {pages} {125126} (\bibinfo {year} {2019})}\BibitemShut {NoStop}%
\bibitem [{\citenamefont {Zhang}\ \emph {et~al.}(2022)\citenamefont {Zhang},
  \citenamefont {Yan}, \citenamefont {Wu}, \citenamefont {He}, \citenamefont
  {Zhang}, \citenamefont {Ni}, \citenamefont {Jiang}, \citenamefont {Qin},
  \citenamefont {Jin}, \citenamefont {Yuan}, \citenamefont {Zhu}, \citenamefont
  {Chen}, \citenamefont {Zhou}, \citenamefont {Li}, \citenamefont {Luo},\ and\
  \citenamefont {Jin}}]{ZhangWPRamanTheory}%
  \BibitemOpen
  \bibfield  {author} {\bibinfo {author} {\bibfnamefont {Y.}~\bibnamefont
  {Zhang}}, \bibinfo {author} {\bibfnamefont {L.}~\bibnamefont {Yan}}, \bibinfo
  {author} {\bibfnamefont {W.}~\bibnamefont {Wu}}, \bibinfo {author}
  {\bibfnamefont {G.}~\bibnamefont {He}}, \bibinfo {author} {\bibfnamefont
  {J.}~\bibnamefont {Zhang}}, \bibinfo {author} {\bibfnamefont
  {Z.}~\bibnamefont {Ni}}, \bibinfo {author} {\bibfnamefont {X.}~\bibnamefont
  {Jiang}}, \bibinfo {author} {\bibfnamefont {M.}~\bibnamefont {Qin}}, \bibinfo
  {author} {\bibfnamefont {F.}~\bibnamefont {Jin}}, \bibinfo {author}
  {\bibfnamefont {J.}~\bibnamefont {Yuan}}, \bibinfo {author} {\bibfnamefont
  {B.}~\bibnamefont {Zhu}}, \bibinfo {author} {\bibfnamefont {Q.}~\bibnamefont
  {Chen}}, \bibinfo {author} {\bibfnamefont {L.}~\bibnamefont {Zhou}}, \bibinfo
  {author} {\bibfnamefont {Y.}~\bibnamefont {Li}}, \bibinfo {author}
  {\bibfnamefont {J.}~\bibnamefont {Luo}}, \ and\ \bibinfo {author}
  {\bibfnamefont {K.}~\bibnamefont {Jin}},\ }\href {\doibase
  10.1103/PhysRevB.105.174511} {\bibfield  {journal} {\bibinfo  {journal}
  {Phys. Rev. B}\ }\textbf {\bibinfo {volume} {105}},\ \bibinfo {pages}
  {174511} (\bibinfo {year} {2022})}\BibitemShut {NoStop}%
\bibitem [{\citenamefont {Lenz}\ and\ \citenamefont
  {Gruehn}(1997)}]{LenzCVTDevelopments}%
  \BibitemOpen
  \bibfield  {author} {\bibinfo {author} {\bibfnamefont {M.}~\bibnamefont
  {Lenz}}\ and\ \bibinfo {author} {\bibfnamefont {R.}~\bibnamefont {Gruehn}},\
  }\href@noop {} {\bibfield  {journal} {\bibinfo  {journal} {Chem. Rev.}\
  }\textbf {\bibinfo {volume} {97}},\ \bibinfo {pages} {2967} (\bibinfo {year}
  {1997})}\BibitemShut {NoStop}%
\bibitem [{\citenamefont {Martin}\ and\ \citenamefont
  {Gruehn}(1990)}]{MartinOxidesCVT}%
  \BibitemOpen
  \bibfield  {author} {\bibinfo {author} {\bibfnamefont {J.}~\bibnamefont
  {Martin}}\ and\ \bibinfo {author} {\bibfnamefont {R.}~\bibnamefont
  {Gruehn}},\ }\href {\doibase https://doi.org/10.1016/0167-2738(90)90465-4}
  {\bibfield  {journal} {\bibinfo  {journal} {Solid State Ionics}\ }\textbf
  {\bibinfo {volume} {43}},\ \bibinfo {pages} {19 } (\bibinfo {year}
  {1990})}\BibitemShut {NoStop}%
\bibitem [{\citenamefont {Mathis}\ \emph {et~al.}(1991)\citenamefont {Mathis},
  \citenamefont {Glaum},\ and\ \citenamefont {Gruehn}}]{MathisWO3PReduction}%
  \BibitemOpen
  \bibfield  {author} {\bibinfo {author} {\bibfnamefont {H.}~\bibnamefont
  {Mathis}}, \bibinfo {author} {\bibfnamefont {R.}~\bibnamefont {Glaum}}, \
  and\ \bibinfo {author} {\bibfnamefont {R.}~\bibnamefont {Gruehn}},\ }\href
  {\doibase 10.3891/acta.chem.scand.45-0781} {\bibfield  {journal} {\bibinfo
  {journal} {Acta Chem. Scand.}\ }\textbf {\bibinfo {volume} {45}},\ \bibinfo
  {pages} {781} (\bibinfo {year} {1991})}\BibitemShut {NoStop}%
\bibitem [{\citenamefont {Sch\"onemann}\ \emph {et~al.}(2017)\citenamefont
  {Sch\"onemann}, \citenamefont {Aryal}, \citenamefont {Zhou}, \citenamefont
  {Chiu}, \citenamefont {Chen}, \citenamefont {Martin}, \citenamefont
  {McCandless}, \citenamefont {Chan}, \citenamefont {Manousakis},\ and\
  \citenamefont {Balicas}}]{SchoenemannWP2}%
  \BibitemOpen
  \bibfield  {author} {\bibinfo {author} {\bibfnamefont {R.}~\bibnamefont
  {Sch\"onemann}}, \bibinfo {author} {\bibfnamefont {N.}~\bibnamefont {Aryal}},
  \bibinfo {author} {\bibfnamefont {Q.}~\bibnamefont {Zhou}}, \bibinfo {author}
  {\bibfnamefont {Y.-C.}\ \bibnamefont {Chiu}}, \bibinfo {author}
  {\bibfnamefont {K.-W.}\ \bibnamefont {Chen}}, \bibinfo {author}
  {\bibfnamefont {T.~J.}\ \bibnamefont {Martin}}, \bibinfo {author}
  {\bibfnamefont {G.~T.}\ \bibnamefont {McCandless}}, \bibinfo {author}
  {\bibfnamefont {J.~Y.}\ \bibnamefont {Chan}}, \bibinfo {author}
  {\bibfnamefont {E.}~\bibnamefont {Manousakis}}, \ and\ \bibinfo {author}
  {\bibfnamefont {L.}~\bibnamefont {Balicas}},\ }\href {\doibase
  10.1103/PhysRevB.96.121108} {\bibfield  {journal} {\bibinfo  {journal} {Phys.
  Rev. B}\ }\textbf {\bibinfo {volume} {96}},\ \bibinfo {pages} {121108(R)}
  (\bibinfo {year} {2017})}\BibitemShut {NoStop}%
\bibitem [{\citenamefont {Xiang}\ \emph {et~al.}(2020)\citenamefont {Xiang},
  \citenamefont {Song}, \citenamefont {Zhou}, \citenamefont {Liang},
  \citenamefont {Xu}, \citenamefont {Qin}, \citenamefont {Wang}, \citenamefont
  {Hong}, \citenamefont {Dai}, \citenamefont {Zhou}, \citenamefont {Liang},
  \citenamefont {Yin}, \citenamefont {Zhao}, \citenamefont {Peng},
  \citenamefont {Yu},\ and\ \citenamefont {Wang}}]{XiangWPGrowth}%
  \BibitemOpen
  \bibfield  {author} {\bibinfo {author} {\bibfnamefont {X.-J.}\ \bibnamefont
  {Xiang}}, \bibinfo {author} {\bibfnamefont {G.-Z.}\ \bibnamefont {Song}},
  \bibinfo {author} {\bibfnamefont {X.-F.}\ \bibnamefont {Zhou}}, \bibinfo
  {author} {\bibfnamefont {H.}~\bibnamefont {Liang}}, \bibinfo {author}
  {\bibfnamefont {Y.}~\bibnamefont {Xu}}, \bibinfo {author} {\bibfnamefont
  {S.-J.}\ \bibnamefont {Qin}}, \bibinfo {author} {\bibfnamefont {J.-P.}\
  \bibnamefont {Wang}}, \bibinfo {author} {\bibfnamefont {F.}~\bibnamefont
  {Hong}}, \bibinfo {author} {\bibfnamefont {J.-H.}\ \bibnamefont {Dai}},
  \bibinfo {author} {\bibfnamefont {B.-W.}\ \bibnamefont {Zhou}}, \bibinfo
  {author} {\bibfnamefont {W.-J.}\ \bibnamefont {Liang}}, \bibinfo {author}
  {\bibfnamefont {Y.-Y.}\ \bibnamefont {Yin}}, \bibinfo {author} {\bibfnamefont
  {Y.-S.}\ \bibnamefont {Zhao}}, \bibinfo {author} {\bibfnamefont
  {F.}~\bibnamefont {Peng}}, \bibinfo {author} {\bibfnamefont {X.-H.}\
  \bibnamefont {Yu}}, \ and\ \bibinfo {author} {\bibfnamefont {S.-M.}\
  \bibnamefont {Wang}},\ }\href {\doibase 10.1088/1674-1056/ab928b} {\bibfield
  {journal} {\bibinfo  {journal} {Chinese Phys. B}\ }\textbf {\bibinfo {volume}
  {29}},\ \bibinfo {pages} {088202} (\bibinfo {year} {2020})}\BibitemShut
  {NoStop}%
\bibitem [{\citenamefont {Tayran}\ and\ \citenamefont
  {Çakmak}(2019)}]{TayranWPTheory}%
  \BibitemOpen
  \bibfield  {author} {\bibinfo {author} {\bibfnamefont {C.}~\bibnamefont
  {Tayran}}\ and\ \bibinfo {author} {\bibfnamefont {M.}~\bibnamefont
  {Çakmak}},\ }\href {\doibase 10.1063/1.5122795} {\bibfield  {journal}
  {\bibinfo  {journal} {J. Appl. Phys.}\ }\textbf {\bibinfo {volume} {126}},\
  \bibinfo {pages} {175103} (\bibinfo {year} {2019})}\BibitemShut {NoStop}%
\bibitem [{\citenamefont {Takase}\ and\ \citenamefont
  {Kasuya}(1980)}]{TakaseMnP}%
  \BibitemOpen
  \bibfield  {author} {\bibinfo {author} {\bibfnamefont {A.}~\bibnamefont
  {Takase}}\ and\ \bibinfo {author} {\bibfnamefont {T.}~\bibnamefont
  {Kasuya}},\ }\href {\doibase 10.1143/JPSJ.48.430} {\bibfield  {journal}
  {\bibinfo  {journal} {J. Phys. Soc. Jpn.}\ }\textbf {\bibinfo {volume}
  {48}},\ \bibinfo {pages} {430} (\bibinfo {year} {1980})}\BibitemShut
  {NoStop}%
\bibitem [{\citenamefont {Ali}\ \emph {et~al.}(2015)\citenamefont {Ali},
  \citenamefont {Schoop}, \citenamefont {Xiong}, \citenamefont {Flynn},
  \citenamefont {Gibson}, \citenamefont {Hirschberger}, \citenamefont {Ong},\
  and\ \citenamefont {Cava}}]{AliWTe2}%
  \BibitemOpen
  \bibfield  {author} {\bibinfo {author} {\bibfnamefont {M.~N.}\ \bibnamefont
  {Ali}}, \bibinfo {author} {\bibfnamefont {L.}~\bibnamefont {Schoop}},
  \bibinfo {author} {\bibfnamefont {J.}~\bibnamefont {Xiong}}, \bibinfo
  {author} {\bibfnamefont {S.}~\bibnamefont {Flynn}}, \bibinfo {author}
  {\bibfnamefont {Q.}~\bibnamefont {Gibson}}, \bibinfo {author} {\bibfnamefont
  {M.}~\bibnamefont {Hirschberger}}, \bibinfo {author} {\bibfnamefont {N.~P.}\
  \bibnamefont {Ong}}, \ and\ \bibinfo {author} {\bibfnamefont {R.~J.}\
  \bibnamefont {Cava}},\ }\href {\doibase 10.1209/0295-5075/110/67002}
  {\bibfield  {journal} {\bibinfo  {journal} {Europhys. Lett.}\ }\textbf
  {\bibinfo {volume} {110}},\ \bibinfo {pages} {67002} (\bibinfo {year}
  {2015})}\BibitemShut {NoStop}%
\bibitem [{\citenamefont {Fallah~Tafti}\ \emph {et~al.}(2016)\citenamefont
  {Fallah~Tafti}, \citenamefont {Gibson}, \citenamefont {Kushwaha},
  \citenamefont {Krizan}, \citenamefont {Haldolaarachchige},\ and\
  \citenamefont {Cava}}]{TaftiXMR}%
  \BibitemOpen
  \bibfield  {author} {\bibinfo {author} {\bibfnamefont {F.}~\bibnamefont
  {Fallah~Tafti}}, \bibinfo {author} {\bibfnamefont {Q.}~\bibnamefont
  {Gibson}}, \bibinfo {author} {\bibfnamefont {S.}~\bibnamefont {Kushwaha}},
  \bibinfo {author} {\bibfnamefont {J.~W.}\ \bibnamefont {Krizan}}, \bibinfo
  {author} {\bibfnamefont {N.}~\bibnamefont {Haldolaarachchige}}, \ and\
  \bibinfo {author} {\bibfnamefont {R.~J.}\ \bibnamefont {Cava}},\ }\href
  {\doibase 10.1073/pnas.1607319113} {\bibfield  {journal} {\bibinfo  {journal}
  {PNAS}\ }\textbf {\bibinfo {volume} {113}},\ \bibinfo {pages} {E3475}
  (\bibinfo {year} {2016})}\BibitemShut {NoStop}%
\bibitem [{\citenamefont {Campbell}\ \emph {et~al.}(2018)\citenamefont
  {Campbell}, \citenamefont {Wang}, \citenamefont {Eckberg}, \citenamefont
  {Graf}, \citenamefont {Hodovanets},\ and\ \citenamefont
  {Paglione}}]{CampbellCoAs}%
  \BibitemOpen
  \bibfield  {author} {\bibinfo {author} {\bibfnamefont {D.~J.}\ \bibnamefont
  {Campbell}}, \bibinfo {author} {\bibfnamefont {L.}~\bibnamefont {Wang}},
  \bibinfo {author} {\bibfnamefont {C.}~\bibnamefont {Eckberg}}, \bibinfo
  {author} {\bibfnamefont {D.}~\bibnamefont {Graf}}, \bibinfo {author}
  {\bibfnamefont {H.}~\bibnamefont {Hodovanets}}, \ and\ \bibinfo {author}
  {\bibfnamefont {J.}~\bibnamefont {Paglione}},\ }\href {\doibase
  10.1103/PhysRevB.97.174410} {\bibfield  {journal} {\bibinfo  {journal} {Phys.
  Rev. B}\ }\textbf {\bibinfo {volume} {97}},\ \bibinfo {pages} {174410}
  (\bibinfo {year} {2018})}\BibitemShut {NoStop}%
\bibitem [{\citenamefont {Segawa}\ and\ \citenamefont
  {Ando}(2009)}]{SegawaFeAs}%
  \BibitemOpen
  \bibfield  {author} {\bibinfo {author} {\bibfnamefont {K.}~\bibnamefont
  {Segawa}}\ and\ \bibinfo {author} {\bibfnamefont {Y.}~\bibnamefont {Ando}},\
  }\href@noop {} {\bibfield  {journal} {\bibinfo  {journal} {J. Phys. Soc.
  Jpn.}\ }\textbf {\bibinfo {volume} {78}},\ \bibinfo {pages} {104720}
  (\bibinfo {year} {2009})}\BibitemShut {NoStop}%
\bibitem [{\citenamefont {Cairns}\ \emph {et~al.}(2020)\citenamefont {Cairns},
  \citenamefont {Stevens}, \citenamefont {D~O’Neill},\ and\ \citenamefont
  {Huxley}}]{CairnsUTe2Variation}%
  \BibitemOpen
  \bibfield  {author} {\bibinfo {author} {\bibfnamefont {L.~P.}\ \bibnamefont
  {Cairns}}, \bibinfo {author} {\bibfnamefont {C.~R.}\ \bibnamefont {Stevens}},
  \bibinfo {author} {\bibfnamefont {C.}~\bibnamefont {D~O’Neill}}, \ and\
  \bibinfo {author} {\bibfnamefont {A.}~\bibnamefont {Huxley}},\ }\href
  {\doibase 10.1088/1361-648X/ab9c5d} {\bibfield  {journal} {\bibinfo
  {journal} {J. Phys.: Condens. Matter}\ }\textbf {\bibinfo {volume} {32}},\
  \bibinfo {pages} {415602} (\bibinfo {year} {2020})}\BibitemShut {NoStop}%
\bibitem [{\citenamefont {Rosa}\ \emph {et~al.}(2022)\citenamefont {Rosa},
  \citenamefont {Weiland}, \citenamefont {Fender}, \citenamefont {Scott},
  \citenamefont {Ronning}, \citenamefont {Thompson}, \citenamefont {Bauer},\
  and\ \citenamefont {Thomas}}]{RosaUTe2Variation}%
  \BibitemOpen
  \bibfield  {author} {\bibinfo {author} {\bibfnamefont {P.~F.}\ \bibnamefont
  {Rosa}}, \bibinfo {author} {\bibfnamefont {A.}~\bibnamefont {Weiland}},
  \bibinfo {author} {\bibfnamefont {S.~S.}\ \bibnamefont {Fender}}, \bibinfo
  {author} {\bibfnamefont {B.~L.}\ \bibnamefont {Scott}}, \bibinfo {author}
  {\bibfnamefont {F.}~\bibnamefont {Ronning}}, \bibinfo {author} {\bibfnamefont
  {J.~D.}\ \bibnamefont {Thompson}}, \bibinfo {author} {\bibfnamefont {E.~D.}\
  \bibnamefont {Bauer}}, \ and\ \bibinfo {author} {\bibfnamefont {S.~M.}\
  \bibnamefont {Thomas}},\ }\href@noop {} {\bibfield  {journal} {\bibinfo
  {journal} {Commun. Mater.}\ }\textbf {\bibinfo {volume} {3}},\ \bibinfo
  {pages} {1} (\bibinfo {year} {2022})}\BibitemShut {NoStop}%
\bibitem [{\citenamefont {Sakai}\ \emph {et~al.}(2022)\citenamefont {Sakai},
  \citenamefont {Opletal}, \citenamefont {Tokiwa}, \citenamefont {Yamamoto},
  \citenamefont {Tokunaga}, \citenamefont {Kambe},\ and\ \citenamefont
  {Haga}}]{SakaiUTe2Growth}%
  \BibitemOpen
  \bibfield  {author} {\bibinfo {author} {\bibfnamefont {H.}~\bibnamefont
  {Sakai}}, \bibinfo {author} {\bibfnamefont {P.}~\bibnamefont {Opletal}},
  \bibinfo {author} {\bibfnamefont {Y.}~\bibnamefont {Tokiwa}}, \bibinfo
  {author} {\bibfnamefont {E.}~\bibnamefont {Yamamoto}}, \bibinfo {author}
  {\bibfnamefont {Y.}~\bibnamefont {Tokunaga}}, \bibinfo {author}
  {\bibfnamefont {S.}~\bibnamefont {Kambe}}, \ and\ \bibinfo {author}
  {\bibfnamefont {Y.}~\bibnamefont {Haga}},\ }\href {\doibase
  10.1103/PhysRevMaterials.6.073401} {\bibfield  {journal} {\bibinfo  {journal}
  {Phys. Rev. Mater.}\ }\textbf {\bibinfo {volume} {6}},\ \bibinfo {pages}
  {073401} (\bibinfo {year} {2022})}\BibitemShut {NoStop}%
\bibitem [{\citenamefont {Cheng}\ and\ \citenamefont
  {Luo}(2017)}]{ChengCrAsMnP}%
  \BibitemOpen
  \bibfield  {author} {\bibinfo {author} {\bibfnamefont {J.}~\bibnamefont
  {Cheng}}\ and\ \bibinfo {author} {\bibfnamefont {J.}~\bibnamefont {Luo}},\
  }\href {\doibase https://doi.org/10.1088/1361-648X/aa7b01} {\bibfield
  {journal} {\bibinfo  {journal} {J. Phys.-Condens. Mat.}\ }\textbf {\bibinfo
  {volume} {29}},\ \bibinfo {pages} {383003} (\bibinfo {year}
  {2017})}\BibitemShut {NoStop}%
\bibitem [{\citenamefont {Guo}\ \emph {et~al.}(2018)\citenamefont {Guo},
  \citenamefont {Smidman}, \citenamefont {Shen}, \citenamefont {Wu},
  \citenamefont {Lin}, \citenamefont {Han}, \citenamefont {Chen}, \citenamefont
  {Wu}, \citenamefont {Wang}, \citenamefont {Jiang}, \citenamefont {Lu},
  \citenamefont {Hu}, \citenamefont {Luo},\ and\ \citenamefont
  {Yuan}}]{GuoCrAsTriplet}%
  \BibitemOpen
  \bibfield  {author} {\bibinfo {author} {\bibfnamefont {C.~Y.}\ \bibnamefont
  {Guo}}, \bibinfo {author} {\bibfnamefont {M.}~\bibnamefont {Smidman}},
  \bibinfo {author} {\bibfnamefont {B.}~\bibnamefont {Shen}}, \bibinfo {author}
  {\bibfnamefont {W.}~\bibnamefont {Wu}}, \bibinfo {author} {\bibfnamefont
  {F.~K.}\ \bibnamefont {Lin}}, \bibinfo {author} {\bibfnamefont {X.~L.}\
  \bibnamefont {Han}}, \bibinfo {author} {\bibfnamefont {Y.}~\bibnamefont
  {Chen}}, \bibinfo {author} {\bibfnamefont {F.}~\bibnamefont {Wu}}, \bibinfo
  {author} {\bibfnamefont {Y.~F.}\ \bibnamefont {Wang}}, \bibinfo {author}
  {\bibfnamefont {W.~B.}\ \bibnamefont {Jiang}}, \bibinfo {author}
  {\bibfnamefont {X.}~\bibnamefont {Lu}}, \bibinfo {author} {\bibfnamefont
  {J.~P.}\ \bibnamefont {Hu}}, \bibinfo {author} {\bibfnamefont {J.~L.}\
  \bibnamefont {Luo}}, \ and\ \bibinfo {author} {\bibfnamefont {H.~Q.}\
  \bibnamefont {Yuan}},\ }\href {\doibase 10.1103/PhysRevB.98.024520}
  {\bibfield  {journal} {\bibinfo  {journal} {Phys. Rev. B}\ }\textbf {\bibinfo
  {volume} {98}},\ \bibinfo {pages} {024520} (\bibinfo {year}
  {2018})}\BibitemShut {NoStop}%
\bibitem [{\citenamefont {Kotegawa}\ \emph {et~al.}(2015)\citenamefont
  {Kotegawa}, \citenamefont {Nakahara}, \citenamefont {Akamatsu}, \citenamefont
  {Tou}, \citenamefont {Sugawara},\ and\ \citenamefont
  {Harima}}]{KotegawaCrAsNQR}%
  \BibitemOpen
  \bibfield  {author} {\bibinfo {author} {\bibfnamefont {H.}~\bibnamefont
  {Kotegawa}}, \bibinfo {author} {\bibfnamefont {S.}~\bibnamefont {Nakahara}},
  \bibinfo {author} {\bibfnamefont {R.}~\bibnamefont {Akamatsu}}, \bibinfo
  {author} {\bibfnamefont {H.}~\bibnamefont {Tou}}, \bibinfo {author}
  {\bibfnamefont {H.}~\bibnamefont {Sugawara}}, \ and\ \bibinfo {author}
  {\bibfnamefont {H.}~\bibnamefont {Harima}},\ }\href {\doibase
  10.1103/PhysRevLett.114.117002} {\bibfield  {journal} {\bibinfo  {journal}
  {Phys. Rev. Lett.}\ }\textbf {\bibinfo {volume} {114}},\ \bibinfo {pages}
  {117002} (\bibinfo {year} {2015})}\BibitemShut {NoStop}%
\bibitem [{\citenamefont {Benhabib}\ \emph {et~al.}(2021)\citenamefont
  {Benhabib}, \citenamefont {Lupien}, \citenamefont {Paul}, \citenamefont
  {Berges}, \citenamefont {Dion}, \citenamefont {Nardone}, \citenamefont
  {Zitouni}, \citenamefont {Mao}, \citenamefont {Maeno}, \citenamefont
  {Georges} \emph {et~al.}}]{BenhabibSr2RuO4}%
  \BibitemOpen
  \bibfield  {author} {\bibinfo {author} {\bibfnamefont {S.}~\bibnamefont
  {Benhabib}}, \bibinfo {author} {\bibfnamefont {C.}~\bibnamefont {Lupien}},
  \bibinfo {author} {\bibfnamefont {I.}~\bibnamefont {Paul}}, \bibinfo {author}
  {\bibfnamefont {L.}~\bibnamefont {Berges}}, \bibinfo {author} {\bibfnamefont
  {M.}~\bibnamefont {Dion}}, \bibinfo {author} {\bibfnamefont {M.}~\bibnamefont
  {Nardone}}, \bibinfo {author} {\bibfnamefont {A.}~\bibnamefont {Zitouni}},
  \bibinfo {author} {\bibfnamefont {Z.}~\bibnamefont {Mao}}, \bibinfo {author}
  {\bibfnamefont {Y.}~\bibnamefont {Maeno}}, \bibinfo {author} {\bibfnamefont
  {A.}~\bibnamefont {Georges}},  \emph {et~al.},\ }\href {\doibase
  10.1038/s41567-020-1033-3} {\bibfield  {journal} {\bibinfo  {journal} {Nat.
  Phys.}\ }\textbf {\bibinfo {volume} {17}},\ \bibinfo {pages} {194} (\bibinfo
  {year} {2021})}\BibitemShut {NoStop}%
\bibitem [{\citenamefont {Ghosh}\ \emph {et~al.}(2021)\citenamefont {Ghosh},
  \citenamefont {Shekhter}, \citenamefont {Jerzembeck}, \citenamefont
  {Kikugawa}, \citenamefont {Sokolov}, \citenamefont {Brando}, \citenamefont
  {Mackenzie}, \citenamefont {Hicks},\ and\ \citenamefont
  {Ramshaw}}]{GhoshSr2RuO4}%
  \BibitemOpen
  \bibfield  {author} {\bibinfo {author} {\bibfnamefont {S.}~\bibnamefont
  {Ghosh}}, \bibinfo {author} {\bibfnamefont {A.}~\bibnamefont {Shekhter}},
  \bibinfo {author} {\bibfnamefont {F.}~\bibnamefont {Jerzembeck}}, \bibinfo
  {author} {\bibfnamefont {N.}~\bibnamefont {Kikugawa}}, \bibinfo {author}
  {\bibfnamefont {D.~A.}\ \bibnamefont {Sokolov}}, \bibinfo {author}
  {\bibfnamefont {M.}~\bibnamefont {Brando}}, \bibinfo {author} {\bibfnamefont
  {A.}~\bibnamefont {Mackenzie}}, \bibinfo {author} {\bibfnamefont {C.~W.}\
  \bibnamefont {Hicks}}, \ and\ \bibinfo {author} {\bibfnamefont
  {B.}~\bibnamefont {Ramshaw}},\ }\href {\doibase 10.1038/s41567-020-1032-4}
  {\bibfield  {journal} {\bibinfo  {journal} {Nat. Phys.}\ }\textbf {\bibinfo
  {volume} {17}},\ \bibinfo {pages} {199} (\bibinfo {year} {2021})}\BibitemShut
  {NoStop}%
\end{thebibliography}%

\end{document}